\documentclass[lettersize,journal]{IEEEtran}
\usepackage{amsmath,amsfonts}
\usepackage{array}
\usepackage[caption=false,font=normalsize,labelfont=sf,textfont=sf]{subfig}
\usepackage{textcomp}
\usepackage{stfloats}
\usepackage{url}
\usepackage{verbatim}
\usepackage{graphicx}
\usepackage{cite}
\usepackage[ruled,linesnumbered]{algorithm2e}
\usepackage{makecell}
\usepackage{booktabs}

\begin{document}

\title{Autonomous and Ubiquitous In-node Learning Algorithms of Active Directed Graphs and Its Storage Behavior}


\author{
    \IEEEauthorblockN{Hui Wei$^{a*}$, Weihua Miao$^{a}$, Fushun Li$^a$}\\
    \IEEEauthorblockA{$^a$ Laboratory of Algorithms for Cognitive Models, School of Computer Science, Fudan University, No.2005 Songhu Road, Shanghai and 200438, China}
}


\markboth{Journal of \LaTeX\ Class Files,~Vol.~14, No.~8, August~2021}%
{Shell \MakeLowercase{\textit{et al.}}: A Sample Article Using IEEEtran.cls for IEEE Journals}


\maketitle

\begin{abstract}
Memory is an important cognitive function for humans. How a brain with such a small power can complete such a complex memory function, the working mechanism behind this is undoubtedly fascinating. Engram theory views memory as the co-activation of specific neuronal clusters. From the perspective of graph theory, nodes represent neurons, and directed edges represent synapses. Then the memory engram is the connected subgraph formed between the activated nodes. In this paper, we use subgraphs as physical carriers of information and propose a parallel distributed information storage algorithm based on node scale in active-directed graphs. An active-directed graph is defined as a graph in which each node has autonomous and independent behavior and relies only on information obtained within the local field of view to make decisions. Unlike static-directed graphs used for recording facts, active-directed graphs are decentralized like biological neuron networks and do not have a super manager who has a global view and can control the behavior of each node. Distinct from traditional algorithms with a global field of view, this algorithm is characterized by nodes collaborating globally on resource usage through their limited local field of view. While this strategy may not achieve global optimality as well as algorithms with a global field of view, it offers better robustness, concurrency, decentralization, and bio-viability. Finally, it was tested in network capacity, fault tolerance, and robustness. It was found that the algorithm exhibits a larger network capacity in a more sparse network structure because the subgraph generated by a single sample is not a whole but consists of multiple weakly connected components. In this case, the network capacity can be understood as the number of permutations of several weakly connected components in the network. The algorithm maintains high recall accuracy and completeness when facing error-containing sample inputs or in the presence of node corruption within the network.
\end{abstract}

\begin{IEEEkeywords}
Directed Graph Storage, Connected Subgraphs, Decentralization.
\end{IEEEkeywords}

\section{Introduction}
\IEEEPARstart{T}{he} traditional theoretical memory models based on psychology and cognitive science are the result of the summary and analysis of many experimental phenomena. It decomposes memory activities into the collaboration and interaction of several functional components, which are usually highly abstract and lack the underlying implementation details. However, most of the computational memory models based on artificial neuron networks, represented by Hopfield network \cite{hopfieldNeuralNetworksPhysical1982} and BAM network \cite{koskoBidirectionalAssociativeMemories1988}, lack biological authenticity. Therefore, it is necessary to establish an intermediate theoretical model that satisfies biological constraints between the macroscopic models of psychology and the molecular mechanism of memory in neurobiology, so as to help us understand the working mechanism of memory.

Marr \cite{marr_vision_2010} proposed a hierarchical framework for analyzing information processing systems, which involves three levels: the theory of computation, representations and algorithms, and hardware implementation. Based on this framework, we analyzed the memory system of the brain. The theory of computation needs to specify the specific requirements of the task, i.e., how the memory storage and retrieval functions are performed by the brain. In the second level, it is necessary to specify the form of memory representation and design algorithms for storing and retrieving memories. The last level is the detailed architecture of connections and communication between neurons in the brain, and since neurobiological studies in this area are more detailed, this paper focuses only on the first two levels. Psychology and cognitive science define and classify memory from a macroscopic point of view, it gives a rough division of memory stages \cite{zlotnik_memory_2019}. While neurobiology shows how neurons connect and communicate with each other from a microscopic perspective. It gives the details of implementing the memory system in the hardware dimension \cite{luoArchitecturesNeuronalCircuits2021a}. However, these perspectives alone are insufficient for a comprehensive understanding of the memory process as they lack detailed representations and algorithms. To gain a more complete understanding of memory, it is necessary to establish reasonable representations and algorithms at an intermediate level. Computer memory, hard disk, and brain memory have similar functions, and their implementation details are compared in Table \ref{tab:new:1}. From the comparison, it can be found that the implementation process of computer storage is very transparent, but there are still systematic gaps in the implementation of human brain memory. It is difficult to provide a logically consistent, sufficiently detailed, and complete theoretical understanding of the entire memory system.

\begin{table*}[]

    \caption{Comparison of computer and brain function implementation details}
    \label{tab:new:1}
    \resizebox{\textwidth}{!}{

\begin{tabular}{p{2cm}p{6.5cm}p{6.5cm}}
\toprule

\textbf{Function implementation details} & \multicolumn{1}{c}{\textbf{Computer}}                                                                                                                                                   & \multicolumn{1}{c}{\textbf{Brain}}                                                                                                                                                                                                                                                \\
\midrule

\textbf{Minimum functional units}        & Data is stored in binary format, such as using the polarity of magnetic particles on a disk to represent 0 and 1.                                                                       & Neuronal membrane potential has two types: resting potential and action potential. Neurons typically maintain a resting potential, but they generate an action potential when activated.                                                                                             \\
\textbf{Rapid memory devices}            & Memory is used to temporarily store the data for CPU calculations and the data for interaction with external devices.                                                                   & Memory is usually stored in a distributed form in multiple brain regions such as the hippocampus and neocortex. From the neural circuit perspective, we know little about how these functional regions collaborate with each other to achieve information localization and response. \\
\textbf{Memory Management}               & Memory is usually managed in a segment-page approach. The operating system uses segments to distinguish different logical information and fixed-size pages to store the actual content. & We have very limited knowledge about how individual memory items are stored in neural circuits and how multiple memory items are allocated to different neuronal groups.  \\

\bottomrule

\end{tabular}
}
\end{table*}

Directed graphs are a classical data structure used in computer science to model the connections between information, such as citation relationships between papers, interaction relationships between proteins, and acquaintance relationships between people in social networks. In recent years, graph databases \cite{francisCypherEvolvingQuery2018} and knowledge graphs \cite{jiSurveyKnowledgeGraphs2022} have been developed to store information in graphs. These databases avoid the need for multi-table unions during traditional relational database queries, providing a more efficient way to access and manage data. However, the directed graphs mentioned above are static and serve as a record of factual information and relationships or a graphical representation of a set of first-order logic propositions. These graphs do not exhibit dynamic behavior independently but act as a data structure required to implement global view-based algorithms. For instance, consider the classical Dijkstra's shortest path algorithm \cite{dijkstraNoteTwoProblems1959}. In each update, a node in the queue that satisfies the conditions is greedily selected. However, this selection operation is not a behavior of the nodes themselves but rather a task of a super manager with a global view. This manager can access information about all nodes, such as the adjacency matrix, enabling it to select the optimal node. In this case, the directed graph is just an information carrier to ensure the algorithm can run efficiently and correctly.

Looking at directed graphs from the perspective of a Multi-Agent System \cite{dorriMultiAgentSystemsSurvey2018}, each node in the graph can be viewed as an independent and fully autonomous agent. Its field of view is limited to the upstream and downstream nodes connected to it, and information is exchanged only between neighboring nodes via directed edges. Every decision made by a node is based on local information, and it can continuously learn and optimize. Now, the directed graph is no longer static but a dynamic system in which neighboring nodes can influence each other. The intrinsic behavior of each node and the external upstream and downstream connections are unique, leading to a rich and complex dynamic behavior of the whole directed graph. In this paper, we term such a directed graph an active-directed graph, which is no longer a manipulated data structure but a cluster of agents with parallel distributed behaviors. An example of an active-directed graph is a biological neural network, where neurons receive and integrate stimulus signals from upstream neurons via dendrites, and subsequently transmit stimulus signals to downstream neurons via axons. Neurons can self-regulate through synaptic plasticity mechanisms like spike-timing-dependent plasticity(STDP) \cite{melizaReceptivefieldModificationRat2006}. Since there is no super-manager with a global view to control the behavior of each neuron or node, the behavior of such active-directed graphs is much more complex than static-directed graphs that only record facts.

A directed graph consists of many nodes and directed edges that can be arranged to form numerous connected subgraphs. These connected subgraphs can be seen as a resource, where a subgraph characterizes a state in which several nodes cooperate or relate to each other. Therefore, the entire directed graph can store or remember much content. In a static-directed graph, the number of connected subgraphs obtained by a path-walking algorithm based on a global view is predictable because there is no uncertainty. However, in an active-directed graph, each node has autonomous behavior and relies only on the information obtained within the local field of view for decision-making. This leads to unpredictable functional connected subgraphs' structure and number. Moreover, since each node can perform incremental adaptive learning, the number of functional subgraphs in the active-directed graph may be much larger and more diverse. To achieve this, corresponding behavior criteria for nodes must be set, which is not required in static-directed graphs.

In a vast active-directed graph, the challenge lies in how to form functional connected subgraphs, consolidate them, efficiently use limited node and path resources, and make multiple connected subgraphs compatible or less interfering with each other. In short, we aim to study how to achieve storage in an active-directed graph. The contribution of this paper is to propose a node-scale-based parallel distributed storage algorithm in active-directed graphs. Unlike traditional algorithms with a global view, the design challenge of such algorithms is that nodes have to collaborate globally on resource usage through their very limited local field of view. This strategy may not achieve global optimality compared to algorithms with global views, but it offers better robustness, concurrency, decentralization, and bio-viability.

\section{Related Work}
Directed graphs have various application modes in storage, and one common mode is the abstract modeling of neuronal networks. One such example is the Hopfield network proposed by John Hopfield in 1982 \cite{hopfieldNeuralNetworksPhysical1982}. It is a fully connected binary recurrent neural network that characterizes the network state by an energy function. Each iteration of the network proceeds towards energy reduction until it reaches a steady state, also known as an attractor. The number of attractors represents the network capacity, which is approximately 0.14N, where N represents the number of nodes in the network. When implementing the associative memory function, the Hopfield network enables complete content retrieval by only part of the sample. However, the capacity of the Hopfield network increases linearly with its network size, making it difficult to preserve too many samples. In 2016, Krotov and Hopfield introduced the discrete modern Hopfield network \cite{krotovDenseAssociativeMemory2016}, which allows network capacity to be extended by changing the network energy function and the update rule, but at the corresponding cost of requiring a large number of hidden layer nodes. Demircigil et al. \cite{demircigilModelAssociativeMemory2017} further extended the energy function by introducing exponential interaction functions, increasing the network capacity. In 2021, Ramsauer, Hubert, et al. \cite{ramsauerHopfieldNetworksAll2021} extended the energy function of modern Hopfield networks from discrete to continuous states while maintaining exponential storage capacity and fast convergence. Hopfield networks, as classical self-associative computational models, enable mapping between vectors of the same dimension. The bidirectional associative memory(BAM) model proposed by Bart Kosko in 1988 \cite{koskoBidirectionalAssociativeMemories1988} can realize both self-association and hetero-association, i.e., mapping between vectors of different dimensions. The model comprises two layers of neurons connected by a weight matrix, which encodes the mapping relationships of all samples. Activating any layer of neurons and iterating through the network results in a correlated output in the neuron on the other layer. In 2021, Bart Kosko \cite{koskoBidirectionalAssociativeMemories2021} introduced a bidirectional backpropagation algorithm in BAM to update matrix parameters dynamically. The original structure was also extended to have any number of hidden layers. The network capacity is increased. In this mode, directed graphs simulate the structure of biological neuron networks, and the network structure is often fixed, such as fully connected or hierarchically connected. The impact of structural parameters such as connectivity, clustering coefficient, and average path length on network performance is not considered. The implementation of the storage function relies more on the weight parameters and update rules of the network, which remains the weight-centric theory. Moreover, all storage contents need to be determined in advance, and the weights are calculated and written at once, making them hard to update incrementally or partially. The local damage to the network will affect the global network, and the scalability of the network scale is not good.

Directed graphs are also commonly used to represent facts and relationships. One popular use case is in graph databases, which employ a graph structure for information queries \cite{anglesSurveyGraphDatabase2008}. In graph databases, nodes represent entities such as people, accounts, or other items, while edges represent connections, such as a friendship relationship. Compared to traditional relational databases, graph databases can perform complex queries more efficiently, as they do not require table join operations but instead search the graph directly. Therefore, they usually offer better performance, especially in the era of big data, where data organization and query complexity have become increasingly important. Graph databases are widely used in various fields, such as in the biomedical domain, for modeling proteins, metabolites, and their relationships, such as digestion and catalysis \cite{thieleCommunitydrivenGlobalReconstruction2013, vastrikReactomeKnowledgeBase2007}. In knowledge graphs \cite{hoganaidanKnowledgeGraphs2021}, information is recorded as RDF triples and stored as graphs, laying the groundwork for achieving goals like semantic search and knowledge inference. In these applications, the role of the directed graph is to record facts, and there is no dynamic behavior. This approach does not address capacity issues or the impact of network structure on storage performance.

In contrast to artificial neural networks that rely on fixed connection patterns, weight parameters, and update rules, or static directed graphs that primarily function for fact recording, this paper presents a method for storing information in a directed network in a distributed manner. This approach leverages the autonomous and dynamic behaviors of numerous nodes in the network, such as resource acquisition and competition. The information content is differentiated based on the distinct combinations of nodes and edges. Subgraphs serve as the information storage carriers without relying on super nodes. Additionally, there is no need for a global view. Information is stored through nodes and edges' local, limited, and adaptive dynamic behaviors. This subgraph-based computational storage model ensures that the stored information remains stable, distinguishable, and fault-tolerant. It also enables incremental storage of information. The performance of this storage method relies on the network's structural characteristics and the nodes' adaptive learning algorithm.

\section{Subgraph-based storage implementation}

In the early 20th century, Richard Semon proposed the concept of engrams, which represents the neural substrate for memory function \cite{josselyn_memory_2020}. He believed that an engram is eventually formed when a group of neurons experiences a persistent physical or chemical change in response to an external stimulus. Subsequently, when the original external stimulus comes again, these engram cells are reactivated, enabling the retrieval of memories. If we try to understand it from the perspective of a directed graph. Nodes represent neurons and directed edges represent synaptic connections between neurons. Then the engram is the subgraph formed between the activated nodes, and the difference in the structure of the subgraph represents the different information contents.

Let $G=(V,E)$ be a directed graph, where $V$ denotes the set of nodes and $E$ denotes the set of edges. If $G'=(V',E')$ is a subgraph of $G$, then it follows that $V'\subseteq V$ and $E'\subseteq E$, denoted as $G'\subseteq G$. In a subgraph-based storage implementation, information is recorded in the form of a series of active nodes and interconnected pathways between them, i.e., a subgraph in the network. For instance, consider the message to be stored as: "While observing a red apple on a tree, I also saw a chirping robin." In this case, nodes representing semantic elements like red, circle, branch, and chirping are activated simultaneously and propagate stimulus along the directed edges in the network. These nodes are referred to as the initial nodes of this sample. During the stimulus propagation process, the initial nodes activate some otherwise inactive nodes, called communication nodes, which are crucial for establishing the pathways. Not all nodes are directly connected by edges in non-fully connected networks, so communication nodes serve as bridges to establish pathways between the initial nodes. These pathways represent associations between semantic elements, such as recalling a robin when seeing an apple again. This occurs when the initial node representing the apple is activated, and the stimulus is passed along the stored pathway in the network, finally activating the node representing the robin. This implementation draws inspiration from cognitive psychology studies on long-term memory \cite{abdouSynapsespecificRepresentationIdentity2018, ohkawaArtificialAssociationPrestored2015}. 

\begin{figure}[h]
  \centering
  \includegraphics[width=\linewidth]{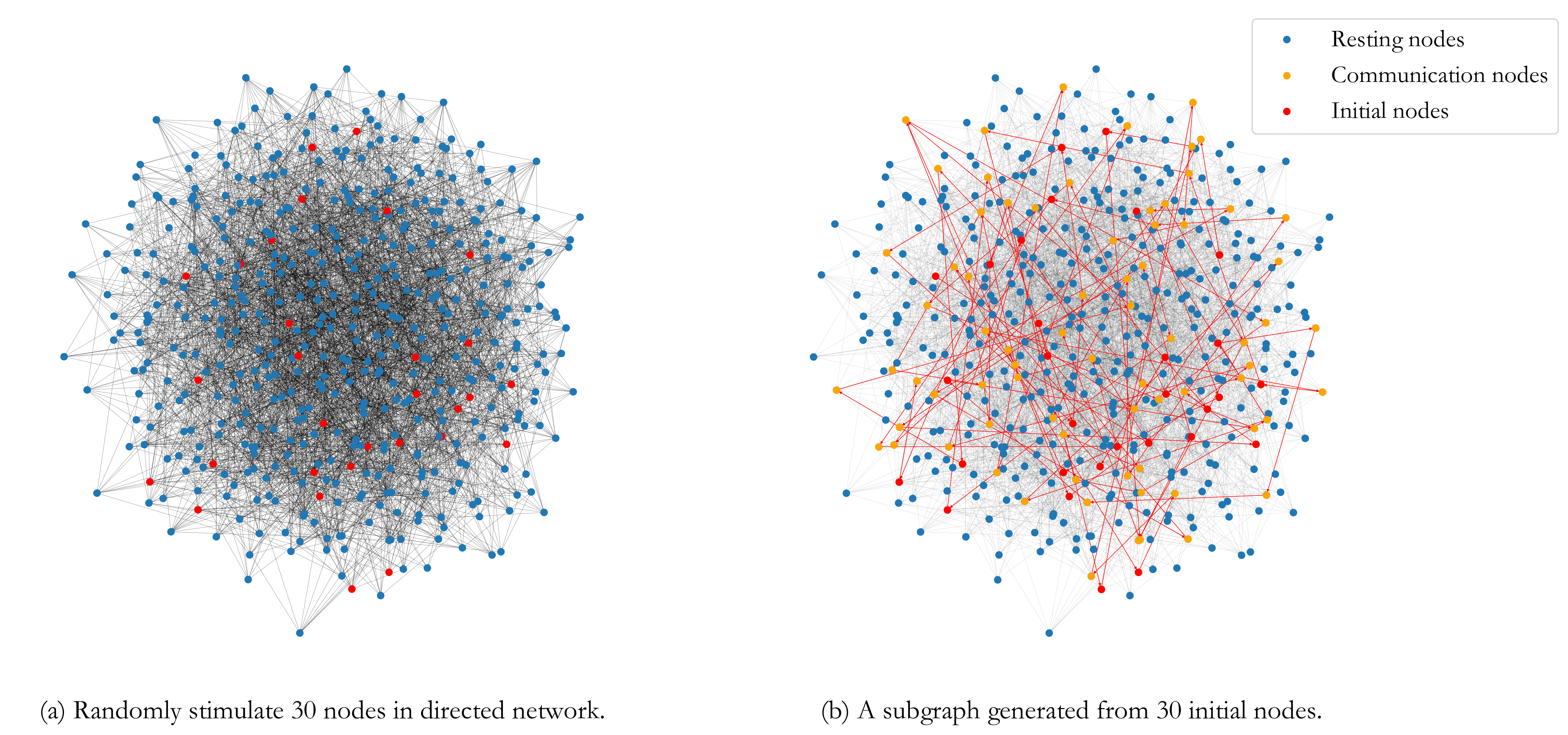}
  \caption{A sample can be stored as a subgraph in a directed graph. (a) Random activation of 30 nodes in the network. (b) Activated nodes propagate stimulus in the network to form a subgraph.}
  \label{fig:1}
\end{figure}

However, the initial idea is insufficient and requires the design of specific implementation details. For example, a node may only characterize a fundamental physical feature, necessitating multiple nodes to represent a concept like an apple. The initial node may activate some communication nodes, which may activate others. Figure \ref{fig:1}a shows 30 randomly selected nodes as initial nodes in a directed network. Figure \ref{fig:1}b shows a stable subgraph obtained by propagating the stimulus of these 30 initial nodes through the network with continuous iterations.

Another factor that makes subgraphs suitable for information storage is the vast number of potential subgraphs present in the network. Given $m=|E|$, there can be up to $2^m$ subgraphs in graph $G$. Thus, using subgraphs as information storage carriers is a promising idea. The challenge lies in ensuring that the subgraphs do not interfere with or confuse each other, effectively utilizing the node and edge resources of the entire directed graph, enabling incremental storage, reducing unfair resource occupation due to varying sample upload orders, and sharing resources among multiple samples. These technical aspects need to be solved by the parallel distributed network adaptive learning algorithm.

\section{Subgraph generation, storage, and retrieval}

The storage and retrieval of samples are processes of the initial nodes propagating the stimulus in the network and eventually forming a stable subgraph. Assuming the stimulus propagation time between nodes is constant. The subgraph eventually reaches a stable state through iterations. The formal definition of a stable subgraph is as follows: let $V_t$ represent the set of all active nodes in the network at time $t$. There exists a minimum time $t'$ such that $V_{t'}\neq V_{t'-1}$ and $V_{t'}=V_{t'+k}$ for any positive integer $k$. At this point, the network is considered to be in a stable state at time $t'$, and the subgraph comprising all active nodes and edges is the stable subgraph. There are two primary concerns: first, how the subgraph is recorded in the network, and second, what rules nodes use for stimulus propagation. Section 4.1 describes the recording of subgraphs, while sections 4.2 to 4.4 outline the stimulus propagation rules. Section 4.5 presents the specific procedure for sample storage and retrieval.

\subsection{The node internal index table records the local upstream and downstream connectivity traces}

The storage of subgraph structures involves recording the connectivity paths between active nodes, which can only be accomplished by the nodes themselves based on their local perspectives. This necessitates that active nodes individually record activation information upstream and downstream of themselves. Define the activation trace of node $u$ as a path fragment consisting of active fan-in nodes and active fan-out nodes of node $u$. Storing the activation traces of all active nodes during this sample storage process will complete the subgraph storage.

\begin{figure}[h]
  \centering
  \includegraphics[width=\linewidth]{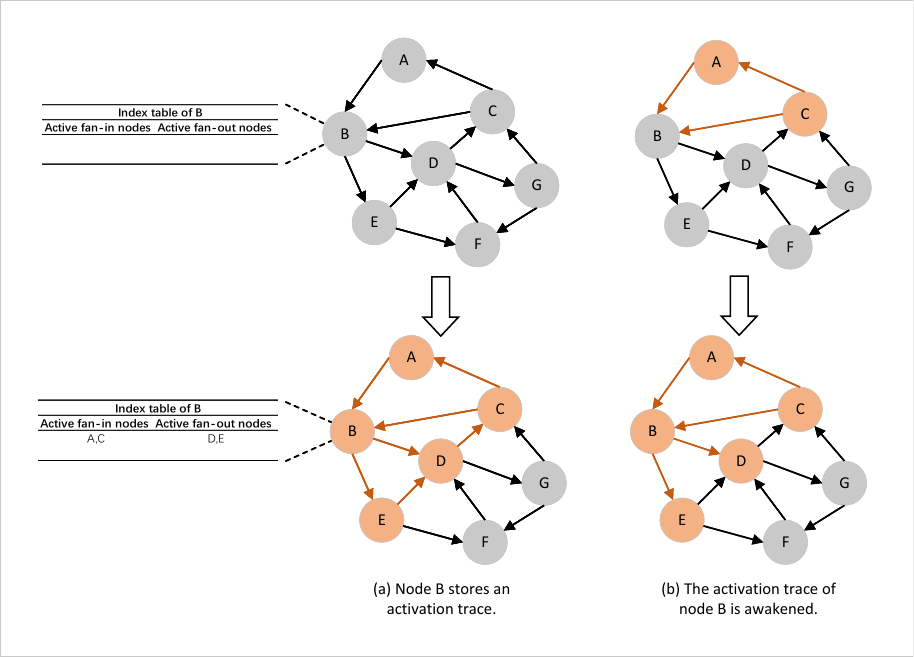}
  \caption{How to record activation traces within a single node. (a) After storing a sample, node B’s index table adds an activation trace. (b) When node B receives the same or similar input again, it reuses the previously recorded activation traces.}
  \label{fig:2}
\end{figure}

In this paper, we store node activation traces by introducing an index table in each node, a data structure with small capacity, easy access, and simple updating. Figure \ref{fig:2} shows the structure of the index table, which contains two columns: the first for active fan-in nodes and the second for active fan-out nodes. Figure \ref{fig:2}a shows that node B's index table contains no content before storing the samples. Once a sample is stored, its corresponding activation trace is saved in its index table. As shown in Figure \ref{fig:2}b, during sample retrieval, if node B receives the same or similar input as the recorded activation traces, it generates the corresponding output based on the historical records in the index table.

Defining an index table inside each node that records upstream and downstream active path pairing relationships may appear straightforward and crude, but it is also biologically feasible. Biological neurons have dendrites that receive inputs from multiple directions and axonal that transmit outputs in different directions, creating a many-to-many connection. Actual physical signaling between upstream and downstream neurons relies on synapses, regulated by combinations of diverse neurotransmitters and ion pumps. These mechanisms precisely control the direction and intensity of positive and negative charge flow. Additionally, differences in synapse location, such as being distal or proximal to the axonal, on dendrites or axons, or on the main pathway or terminal, can precisely control the activation and deactivation of specific action potential transmission pathways. In conclusion, this highly precise and diverse molecular-level and subcellular-level modulation and their combinations equip biological neurons with various pathway control mechanisms at the microcircuit level \cite{tritschMechanismsFunctionsGABA2016, chenRoleIntrinsicExcitability2020}. As a result, a biological neuron can achieve diverse pathway control of signaling within its small neighborhood, relying on a complex set of electrochemical processes \cite{luoArchitecturesNeuronalCircuits2021a}. This has inspired the design of directed graph nodes' internal behavior, allowing them to function like network routers capable of differentially leading fan-outs based on fan-in variations. An index table with limited storage space is a simple, functional equivalent implementation.

\subsection{Intra-node stimulus propagation algorithm}

The creation of subgraphs depends on the propagation of stimulus between nodes. Stimulus propagation consists of two aspects: node activation rules, i.e., how nodes are activated, and stimulus propagation rules, i.e., determining the downstream nodes to which the stimulus is propagated. In this paper, the node activation rule employed is a fixed-probability activation model, where a node will be activated with a fixed probability upon receiving input. The node becomes active and begins delivering stimulus to downstream nodes if successfully activated. Two types of stimulus propagation rules are used in this paper: the first reuses similar historical activation traces, and the second employs a weighted random selection algorithm.

The specific process of stimulus propagation among nodes is as follows: each resting state node has a fixed probability of H to be activated after receiving the stimulus. Once a node is activated, if its index table is empty, it randomly selects several downstream nodes with equal probability for stimulus delivery. If the index table is not empty, the similarity between the input and each item in the node index table is calculated first. In this paper, the F1 score \cite{manningIntroductionInformationRetrieval2008} is used as a metric to evaluate the similarity of two node sequences. The F1 score is a statistical measure of the accuracy of a binary classification model, which is the harmonic mean of precision and recall. When comparing similarity, either one of the node sequences can be treated as the predicted value and the other as the actual value, and the corresponding F1 score is calculated. Higher scores indicate greater similarity. The corresponding historical activation traces are reused if the maximum similarity exceeds the threshold. Otherwise, a weighted random selection algorithm is used to select downstream nodes for stimulus delivery.

The weighted random selection algorithm is based on the frequency of the downstream node appearing in all active fan-out nodes in the current node index table. The higher the frequency of occurrence, the lower the chance of being selected. The introduction of this algorithm allows each active node to distribute stimulus evenly, thus maximizing network resource utilization. The algorithm pseudo-code is shown in Algorithm \ref{alg:1}. The value of $H$ affects the subgraph size. The larger $H$ is, the more nodes participate in subgraph formation and the better the connectivity. However, the corresponding cost of network resources is also larger. In this paper, $H$ is set at 60\% for testing.

\begin{algorithm}
    \caption{$NodeStimulusSpreading(u)$ }\label{alg:1}
    \KwData{$Table[u]$: $u$'s index table. $currentIn[u]$: Current input of $u$. $Threshold$: Similarity threshold. $H$: Probability of being activated. $G[u]$: $u$'s adjacency table. $eOutNum$: The expected value of output size. $freq[v]$: The occurrence frequency of $v$ in the index table of $u$.}
    \KwIn{$u$: Current node}
    \KwOut{$uOut$: The fan-out nodes of $u$}
    \If{$randomValue(0,1) > H$}{
        \tcc{Activation failed}
        return NULL\;
    }
    $mxF1 \leftarrow 0$\;
    $uOut \leftarrow NULL$\;
    \For{$item$ in $Table[u]$}{
        \tcc{Traverse the index table and find the item most similar to $currentIn[u]$}
        $f1Score \leftarrow CalculateF1Score(item.in, currentIn[u])$\;
        \If{$f1Score > Threshold$ and $mxF1 < f1Score$}{
            $mxF1 \leftarrow f1Score$\;
            $uOut \leftarrow item.out$\;
        }
    }
    \If{$uOut \neq NULL$}{
        return $uOut$\;
    }
    \For{$v$ in $G[u]$}{
    $freq[v] \leftarrow getFrequencyOfOccurrence(v,Table[u])$\;
    }
    $uOut \leftarrow randomChoose(G[u], freq, eOutNum)$\;
    return $uOut$\;
\end{algorithm}

\subsection{Node Resource Grabbing Rules}

\begin{figure}[h]
  \centering
  \includegraphics[width=\linewidth]{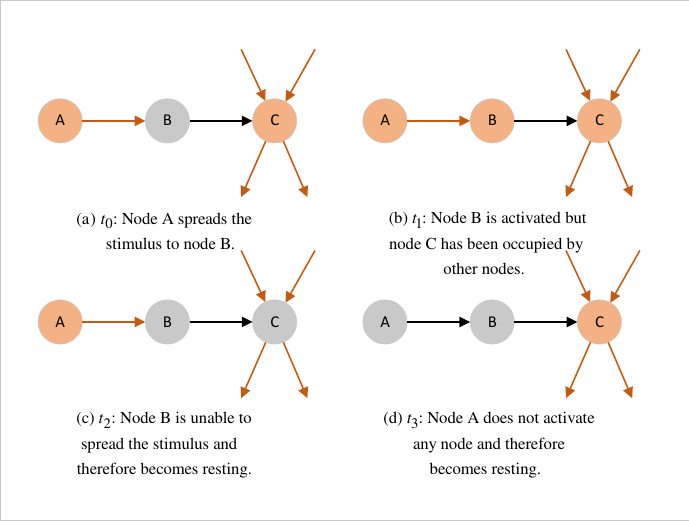}
  \caption{Node resource-grabbing rules. (a) Node A propagates stimulus to node B. (b) After node B is activated, it finds that other nodes have occupied node C. (c) Node B becomes a resting state as it cannot continue to propagate stimulus. (d) Node A becomes a resting state because it has not activated any downstream nodes.}
  \label{fig:3}
\end{figure}

Nodes are considered limited resources in a directed graph, adhering to the first-come, first-served preemption rule. Activating a node can be seen as the occupation of a node resource. When performing stimulus propagation, nodes can acquire the occupancy of downstream nodes to avoid passing stimulus to already occupied nodes. If an active node does not successfully activate any downstream nodes, it returns to a resting state. A change in the state of some active nodes may trigger a chain reaction that causes more active nodes to become resting. This situation is called the avalanche effect. As shown in Figure \ref{fig:3}, at $t_0$, node B receives a stimulus from node A, then subsequently activated at $t_1$. However, because node C has been occupied, node B cannot transmit the stimulus to node C, causing node B to revert to the resting state at the $t_2$ moment. At this point, node A is not activating any nodes due to the change in the state of node B. Therefore, at $t_3$, node A also becomes resting due to the avalanche effect.

\subsection{Several problems are caused by insufficient resources in subgraph generation}

As the number of samples stored in the network increases, new subgraphs may encounter some problems caused by insufficient resources during the generation process, preventing new samples from being stored. The causes of insufficient resources in the network can be broadly classified into four categories:

1.	The node index table has a capacity limit. When the number of samples stored in the network reaches a certain level, it becomes hard to store new samples.

2.	The network is poorly connected, and the stimulus propagation rule carries a certain level of randomness, as well as the existence of the avalanche effect, which leads to an inability to establish a pathway between the initial nodes.

3.	The samples already stored in the network interfere with the samples currently about to be stored. This is because nodes may reuse historical activation traces when they are activated. Although this is an optimization strategy to increase network capacity, it somewhat affects the storage of current samples.

4.	Some active nodes take up too many node resources during the current activation, resulting in no resources available for other nodes.

For the first case, the capacity of the index table can be defined as the maximum number of output types that a node can store, as there may be many different inputs corresponding to the same output. The capacity can also be expanded by reasonably discarding and merging the contents of the index table. Specifically, when a node's index table capacity reaches its upper limit, the node will search for the two most similar activation traces to merge. The similarity here refers to the similarity of the active fan-in nodes in the two activation traces, and merging refers to taking the intersection of the active fan-out nodes of the two activation traces. If the differences between the activation traces are both large, the one with the lowest strength is discarded. The strength here refers to the number of samples that the activation trace has been involved in storing. The more involved, the higher the strength. By reasonably merging and discarding, it is possible to increase network capacity as much as possible at the expense of certain recall accuracy and completeness. The pseudo-code of the algorithm is given by Algorithm \ref{alg:2}.

\begin{algorithm}
    \caption{$ActivationTracesUpdating()$ }\label{alg:2}
    \KwData{$activeNodes$: active nodes set. $Outputs[u]$: A collection of different output in $Table[u]$. $K$: The upper limit of the output type.}
    \For{$u$ in $activePoints$}{
        $Table[u][in[u]] = out[u]$\;
        \If{$out[u]$ not in $Output[u]$}{
            $Output[u].add(out[u])$\;
        }
        \If{$Output[u].size() > K$}{
            \tcc{Find the two most similar outputs to merge}
            $out1, out2 \leftarrow findTwoMostSimilarOutput(Output[u])$\;
            \eIf{$out1 \neq NULL$ and $out2 \neq NULL$}{
                $out3 \leftarrow merge(out1, out2)$\;
                $Output[u].remove(out1)$\;
                $Output[u].remove(out2)$\;
                $Output[u].add(out3)$\;
                \tcc{Change the output of all items whose output is $out1$ or $out2$ in $Table[u]$ to $out3$}
                $changeItemsInTable(Table[u], out1, out3)$\;
                $changeItemsInTable(Table[u], out2, out3)$\;
            }{
                \tcc{If not found, discard the output with the lowest strength}
                $out1 \leftarrow findLowestIntensityOutput(Table[u])$\;
                $Output[u].remove(out1)$\;
                \tcc{Delete all items whose output is $out1$}
                $changeItemsInTable(Table[u], out1, NULL)$\;
            }
        }
    }
\end{algorithm}

\begin{figure}[h]
  \centering
  \includegraphics[width=\linewidth]{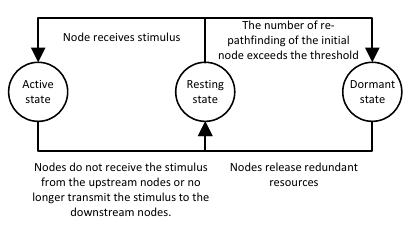}
  \caption{Node state transition graph}
  \label{fig:4}
\end{figure}

\begin{figure}[h]
  \centering
  \includegraphics[scale=0.4]{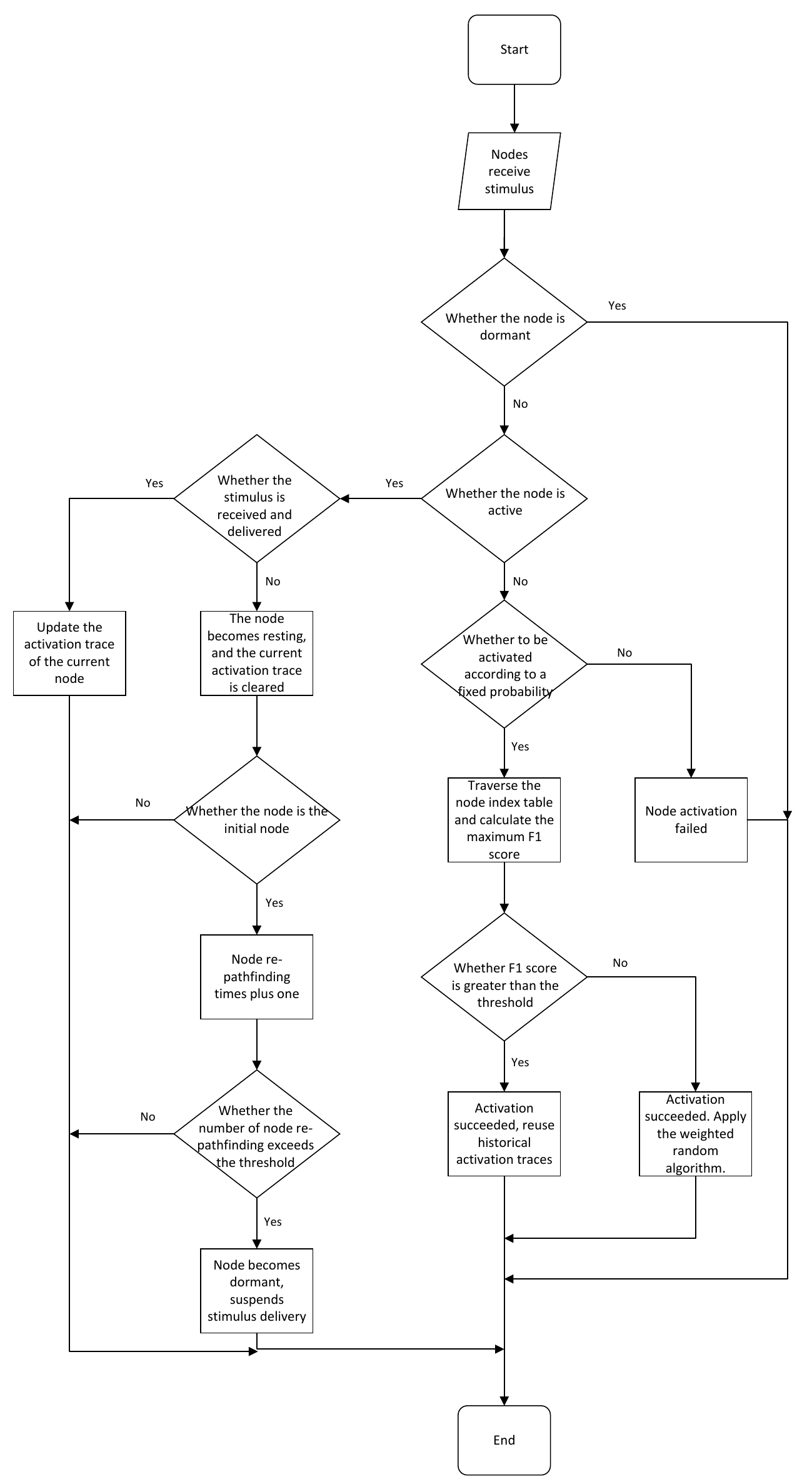}
  \caption{Algorithm flow chart of nodes receiving, processing, and outputting stimulus}
  \label{fig:5}
\end{figure}

The second and third cases can be solved by introducing a re-pathfinding rule. The re-pathfinding rule allows the initial nodes to re-propagate the stimulus to find a path connecting other initial nodes when it does not successfully activate any downstream node. For the fourth case, the active node can release some occupied resources according to its situation. The resource release algorithm is introduced here to solve this problem. When the number of failed re-pathfindings of an initial node reaches a certain threshold, it will enter a dormant state, indicating that it is currently unable to communicate with other nodes. The node in the dormant state suspends pathfinding until the subgraph stabilizes. Active nodes will release some nodes after the subgraph is stabilized to make resources available to dormant nodes. For example, if an active node has three active fan-out nodes, it can actively release the occupation of two of them. After releasing the redundant resources, the nodes in the dormant state will resume pathfinding until the subgraph stabilizes again. If, after releasing the resources, the dormant node is still unable to establish path connections to other active nodes, the network connectivity is considered poor, or there is a conflict between the current sample and the samples already stored in the network. In this case, the subgraph can still be formed. However, there will be some isolated nodes that cannot establish connections with other nodes, leading to a decrease in the subgraph's anti-interference ability and fault tolerance. The pseudo-code of the resource release algorithm is given by Algorithm \ref{alg:3}. Figure \ref{fig:4} shows the transition relationship between the three node states. Figure \ref{fig:5} presents the flowchart of the algorithm from the node's perspective, which includes how nodes process the received stimulus and how downstream nodes are selected for stimulus delivery.

\begin{algorithm}
    \caption{$ResourcesReleasing()$ }\label{alg:3}
    \KwData{$dormantCnt$: The number of nodes in dormant state. $isDormant[u]$:  Whether $u$ is in dormant state. $repathCnt[u]$: Indicates the number of re-pathfinding.}
    \KwOut{$True$: The resource is released successfully and needs to continue to iterate. $False$: No node resources can be released, the iteration is complete.}
    \If{$dormantCnt > 0$}{
        $isReleased \leftarrow False$\;
        \tcc{The resources will be released if the subgraph is stable and several nodes are dormant.}
        \For{$u \leftarrow 0$ to $n$}{
            \If{out degree of $u$ > 1}{
                \tcc{Release redundant resources}
                $isReleased \leftarrow releaseResources(u)$\;
            }
            \If{$isDormant[u] == True$}{
                $isDormant[u] \leftarrow False$\;
                $dormantCnt \leftarrow dormantCnt - 1$\;
                $repathCnt[u] \leftarrow 0$\;
            }
        }
        return $isReleased$\;
    }
    return $False$\;

\end{algorithm}

\subsection{Sample Storage and Retrieval}

\begin{figure}[h]
  \centering
  \includegraphics[scale=0.4]{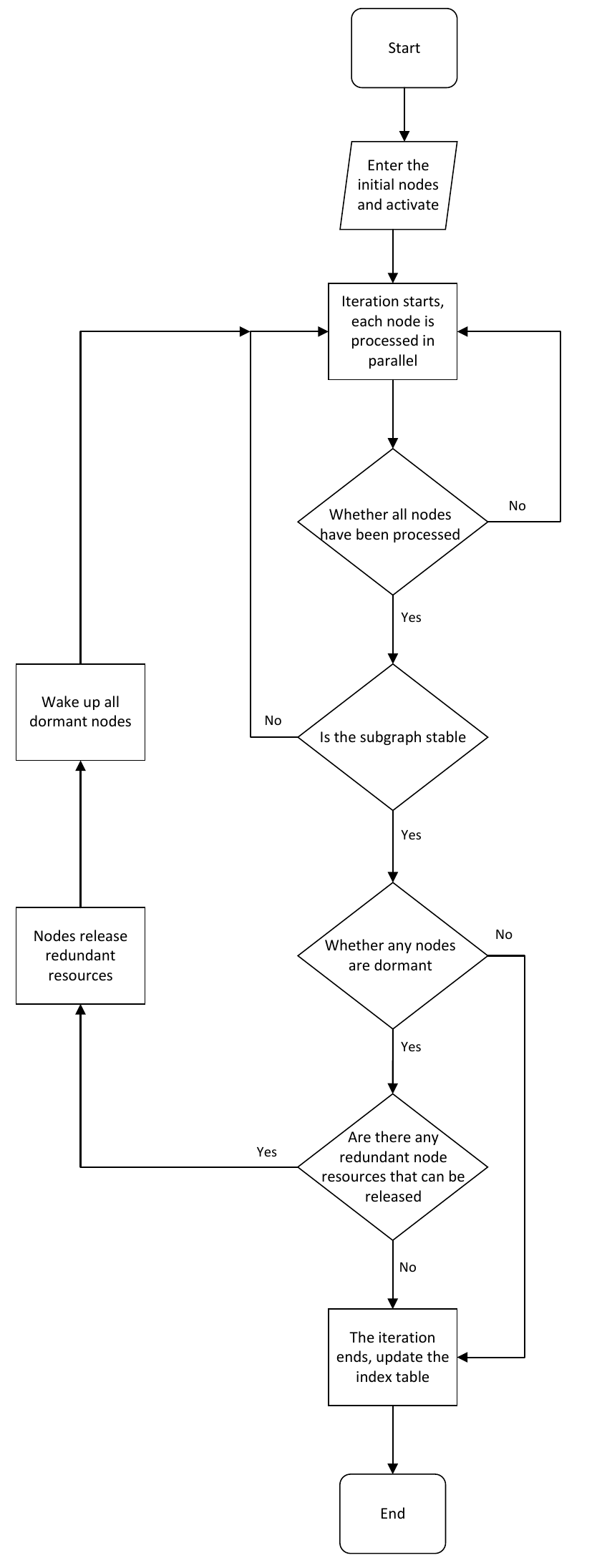}
  \caption{Algorithm flow chart of sample storage}
  \label{fig:6}
\end{figure}

The sample storage process consists of two stages: (1) Stimulus propagation stage: the initial nodes propagate stimulus to other nodes along the directed edges until a stable subgraph is formed. (2) Subgraph consolidation stage: All nodes in the subgraph update their internal index tables, recording the activation traces. Figure \ref{fig:6} shows the storage algorithm from the processor's perspective and the flowchart of the algorithm from the task scheduling perspective. The corresponding pseudo-code is given by Algorithm \ref{alg:4}.

Stimulus propagation stage: Table \ref{tab:1} demonstrates a complete process of generating a stable subgraph through continuous iteration of the initial nodes. At the time $t_0$, the initial nodes are activated, and downstream nodes are chosen for stimulus propagation according to the weighted random selection algorithm. The subsequent $t_1$ and $t_2$ moments represent the continuous stimulus propagation in the network. At the time $t_3$, since downstream nodes B and C of node H have been occupied by other nodes, node H cannot perform stimulus transfer. Therefore, according to the node resource-grabbing rules, node H transitions from the active state to the resting state. At the time $t_4$, downstream node H, excited by node J, reverts to a resting state. At this point, node J does not activate any nodes, and due to the avalanche effect, its state also becomes a resting state. After the end of time $t_4$, node D will no longer propagate stimulus. However, as the initial node, it will follow the re-pathfinding rules, searching for a new path and attempting to participate in the subgraph formation. When re-pathfinding reaches a certain number of attempts, the node will enter a dormant state. Here, it is assumed that node D has entered a dormant state and will halt pathfinding until the subgraph is stable. It can be observed that at time $t_4$, the subgraph is already stable since there will be no change in node states. At this point, it is necessary for other active nodes in the network to release redundant resources, providing node D the opportunity to re-engage in the subgraph formation. At the time $t_5$, node A, which originally occupied both node E and node G resources, can choose to release the occupation of either of the two nodes. Assuming that node E is released, node E will become resting, and the stimulus from node E to node B will also vanish. However, because node B is an initial node, its state will not change. After the resource is released, node D resumes pathfinding and node J is activated at time $t_6$. At the time $t_7$, node H is activated by node J, and the stimulus is passed to the initial node B, forming a path. At this moment, the connected subgraph between active nodes becomes stable, no dormant nodes are present in the network, and the stimulus propagation stage concludes.

\begin{table*}
    \caption{Example of a dynamic process for directed graph stimulus propagation}
    \label{tab:1}
    \resizebox{\textwidth}{!}{
        \begin{tabular}{|c|c|c|}

    \hline
    \makecell*[c]{\textbf{Graph}} & \makecell*[c]{\includegraphics[scale=0.5]{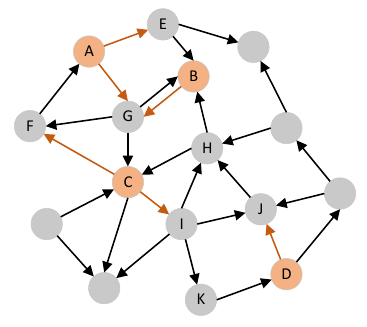}} & \makecell*[c]{\includegraphics[scale=0.5]{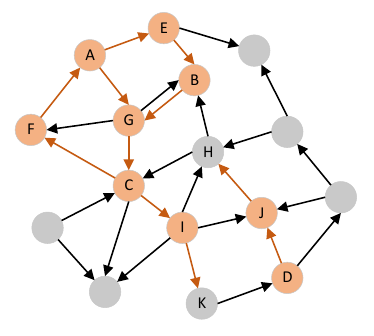}} \\ \hline
    \makecell*[c]{\textbf{Nodes state} \\ \textbf{change}} & \makecell*[c]{$t_0$: Initial nodes \{A,B,C,D\} are activated, \\ which performs stimulus propagation according to \\ a weighted random selection algorithm.} & \makecell*[c]{$t_1$: \{E,F,G,I,J\} becomes active after receiving \\ stimulus from the initial nodes.} \\ \hline

    \makecell*[c]{\textbf{Graph}} & \makecell*[c]{\includegraphics[scale=0.5]{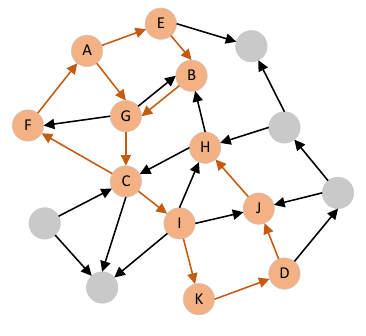}} & \makecell*[c]{\includegraphics[scale=0.5]{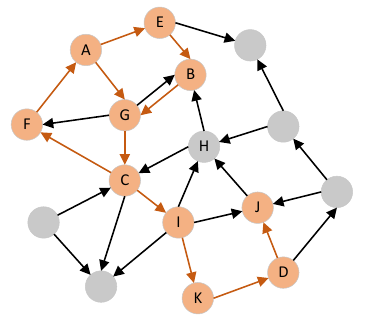}} \\ \hline
    \makecell*[c]{\textbf{Nodes state} \\ \textbf{change}} & \makecell*[c]{$t_2$: \{H,K\} becomes active after receiving \\ stimulus from active nodes.} & \makecell*[c]{$t_3$: The downstream nodes \{B,C\} of H are all occupied, \\ so stimulus cannot be propagated.\\ H becomes resting again.} \\ \hline

    \makecell*[c]{\textbf{Graph}} & \makecell*[c]{\includegraphics[scale=0.5]{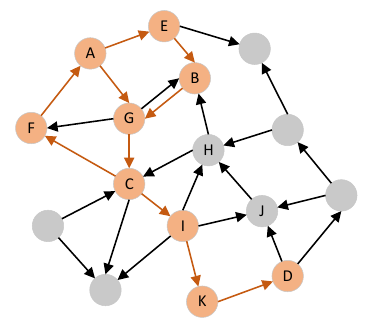}} & \makecell*[c]{\includegraphics[scale=0.5]{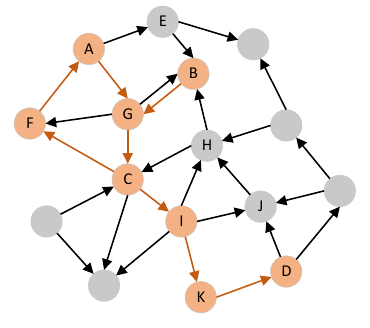}} \\ \hline
    \makecell*[c]{\textbf{Nodes state} \\ \textbf{change}} & \makecell*[c]{$t_4$: After H becomes resting state,\\ according to the avalanche effect, \\ J also becomes resting state.} & \makecell*[c]{$t_5$: The subgraph iterates to a steady state. \\ Start to release resources, \\ and node A releases the occupancy of E.} \\ \hline

    \makecell*[c]{\textbf{Graph}} & \makecell*[c]{\includegraphics[scale=0.5]{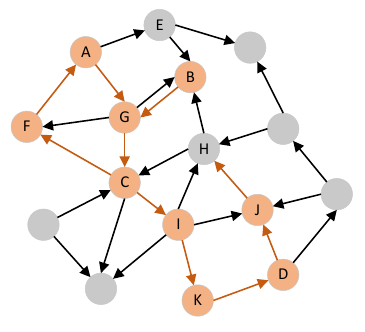}} & \makecell*[c]{\includegraphics[scale=0.5]{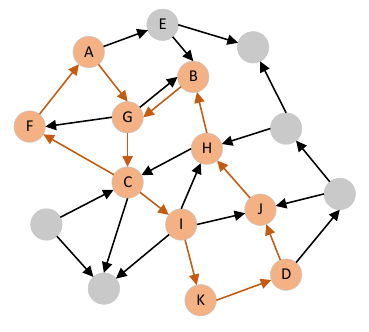}} \\ \hline
    \makecell*[c]{\textbf{Nodes state} \\ \textbf{change}} & \makecell*[c]{$t_6$: After the resources are released,\\ the dormant node D restarts pathfinding \\ and successfully activates J.} & \makecell*[c]{$t_7$: Node H is successfully activated after receiving \\ the stimulus from J and passing the stimulus to B.\\ The subgraph is iterated to a stable state.} \\ \hline

    \end{tabular}}
\end{table*}

Subgraph Consolidation Stage: The primary task of this stage is to store a stable subgraph structure in the network. When the subgraph achieves a stable state, each active node will have corresponding active fan-in and active fan-out nodes, which are activation traces. Storing the subgraph is completed by updating the activation trace of each node in the node's internal index table. The pseudo-code of the index table update algorithm is provided by Algorithm \ref{alg:2}, and the pseudo-code of subgraph generation and preservation is given by Algorithm \ref{alg:4}.

\begin{algorithm}
    \caption{$SubgraphGeneratingAnd\\Saving(initialNodes)$}\label{alg:4}
    \KwData{$isActive[u]$: whether $u$ is activated. $isInitialNode[u]$: Whether $u$ is a initial node. $changed$: Whether the network has iterated to a stable state.}
    \KwIn{$initialNodes$}
    $dormantCnt \leftarrow 0$\;
    \For{$u$ in $initialNodes$}{
        $isActive[u] \leftarrow True$\;
        $isInitialNode[u] \leftarrow True$\;
    }
    \While{$True$}{
        $changed \leftarrow False$\;
        \For{$u \leftarrow 0$ to $n$}{
            \tcc{$u$ receives stimulus from upstream nodes}
            $receiveStimulus(u)$\;
        }
        \For{$u \leftarrow 0$ to $n$}{
            \If{$isDormant[u] == True$}{
                continue\;
            }
            \tcc{$u$ delivers stimulus to downstream nodes}
            $NodeStimulusSpreading(u)$\;
        }
        \tcc{The network is not stable, continue to iterate}
        \If{$changed == True$}{
            continue\;
        }
        \tcc{The network is stable and releases resources}
        \If{$ResourcesReleasing() == False$}{
            break\;
        }
    }
    $ActivationTracesUpdating()$\;
\end{algorithm}

The process of sample retrieval closely resembles sample storage. However, the sample retrieval process is simpler, only including the stimulus propagation stage. During the stimulus propagation stage, it will not be activated when a node receives a stimulus transfer from an upstream node and cannot find a similar entry in its internal index table. The initial nodes will not enter the dormant state, and no nodes will release excessively occupied resources. In summary, the sample retrieval algorithm will not cause any changes to the existing network. However, it will only perform stimulus propagation based on the activation traces stored in the internal index table of the node. Since the sample retrieval algorithm process is very similar to the storage algorithm, only the specified part mentioned above needs to be omitted. Therefore, no separate pseudo-code is provided here.

\section{Experiments}

The experiment is primarily divided into four aspects:

1.	Capacity testing: This aims to investigate the number of samples the network can stably store and the factors influencing network capacity.

2.	Fault tolerance testing: This mainly explores the effect of sample retrieval when the input is incomplete or has noise.

3.	Robustness testing: This mainly explores the effect of sample retrieval when some nodes or edges are damaged.

4.	Performance testing on different classical network structures: This mainly explores the algorithm's performance on various classic network structures.

The network used in the experiment is ER random graphs \cite{erdosRandomGraphs1959}. This classic random network model, proposed by Paul Erdős and Alfréd Rényi in 1959, is defined by having a probability p of connection between any two nodes in the network. Extending this definition to directed graphs, two distinct directional edges can be between any two nodes. The probabilities of these two edges existing are independent of each other, and both are equal to $p$.

\subsection{Capacity testing}

Let $G_i=(V_i,E_i)$ represent the subgraph generated of the $i$th sample, where $V_i$ denotes the set of nodes, and $E_i$ denotes the set of edges. Let $G'_i=(V'_i,E'_i)$ represent the subgraph generated during the retrieval of the $i$th sample. Define the accuracy $P_i=\frac{|E_i \cap E'_i|}{|E'_i|}$. Define the completeness $C_i=\frac{|E_i \cap E'_i|}{|E_i|}$. Define an isolated node as an initial node in the subgraph with both in-degree and out-degree equal to zero. Define the sample representation quality, $Q$, as the percentage of non-isolated nodes relative to the initial nodes. If the number of initial nodes in the sample is $s$, and the number of isolated nodes in the subgraph generated by the sample is $l$, then $Q=\frac{s-l}{s}$. In capacity testing, each stored sample should satisfy a high sample representation quality and high completeness and accuracy during sample retrieval to be considered successfully stored by the network. Based on this, we can define the reliable capacity of the network. Let the reliable capacity, $T$, represent the maximum number of samples that the network can successfully store, and for each stored sample, it satisfies $Q_i>0.9$, $i\in [1,T]$, $\bar P=\frac{\sum_{i=1}^{T}P_i}{T}>0.9$, $\bar C=\frac{\sum_{i=1}^{T}C_i}{T}>0.9$.

Assume the network has $n$ nodes, each node has an internal index table with a capacity of $K$, and each subgraph contains, on average, $s$ initial nodes and $c$ communication nodes. For every subgraph stored in the network, there will be an increase or modification in the internal index table entries of the activated nodes. Considering this as a resource, the total number of resources in the network is $nK$, and each subgraph occupies $s+c$ resources. Without considering resource reuse or optimization measures, the network will consume $s+c$ resources for every stored sample, so the network capacity can be roughly represented as $\frac{nK}{s+c}$. If resource reuse is allowed, the calculation of network capacity becomes more complex. In an extreme case, where the resources occupied by the current subgraph are all reused, the upper bound of the network capacity can be roughly expressed as the combination number $\tbinom{nK}{s+c}$. The network capacity obtained from these two different calculation methods differs greatly. In actual testing, there are many other influencing factors, such as different network connectivity and conflicts between samples. Therefore, a theoretical network capacity analysis is difficult, and a specific analysis should be conducted in conjunction with actual testing situations.

Table \ref{tab:2} shows the performance of sample retrieval after storing 1,000 samples in networks. The node index table size is set to $K=20$. The scale of a single sample refers to the size of the initial node set. As shown in Table \ref{tab:2}, the capacity of sparse graphs is typically larger than that of dense graphs with the same node size. The primary distinction between sparse and dense graphs lies in the number of directed edges within the network, which directly influences network connectivity. This can also be observed from the average number of weakly connected components in the subgraphs presented in Table \ref{tab:2}. For networks with the same node size, the more edges they have, the fewer weakly connected components their subgraphs have on average. Generally, the subgraphs generated by samples are not necessarily connected but are composed of multiple connected components. Weakly connected components (WCCs) are defined as components where undirected edges replace all directed edges, and any two points within the component are reachable from one another. The number of connected components reflects the aggregation of the subgraph. A greater number of connected components indicates a more dispersed subgraph, while a smaller number of connected components signifies a more clustered subgraph.

\begin{table*}[]
    \caption{Different-scale network retrieval performance after storing 1000 samples}
    \label{tab:2}

    \resizebox{\textwidth}{!}{
        \begin{tabular}{cccccccc}
            \toprule

            \makecell*[c]{\textbf{Number of} \\ \textbf{network nodes}}    & \makecell*[c]{\textbf{Number} \\ \textbf{of edges}} & \makecell*[c]{\textbf{Single} \\ \textbf{sample scale}} &  \makecell*[c]{\textbf{Subgraph average} \\ \textbf{number of nodes}} & \makecell*[c]{\textbf{Subgraph average} \\ \textbf{number of edges}}  & \makecell*[c]{\textbf{Subgraph average} \\ \textbf{number of WCCs}}  & \makecell*[c]{\textbf{Average} \\ \textbf{accuracy}} & \makecell*[c]{\textbf{Average} \\ \textbf{completeness}} \\ 

            \midrule

            \makecell*[c]{500}            & \makecell*[c]{3101(Sparse)}      & \makecell*[c]{15}                & \makecell*[c]{24.024}          & \makecell*[c]{15.651}          & \makecell*[c]{3.372}           & \makecell*[c]{99.6\%}         & \makecell*[c]{98.4\%}         \\ 
            \makecell*[c]{500}            & \makecell*[c]{3101(Sparse)}        & \makecell*[c]{60}                & \makecell*[c]{85.343}          & \makecell*[c]{70.763}          & \makecell*[c]{10.312}          & \makecell*[c]{97.5\%}         & \makecell*[c]{94.8\%}         \\ 
            \makecell*[c]{500}            & \makecell*[c]{12606(Dense)}     & \makecell*[c]{15}                & \makecell*[c]{28.407}          & \makecell*[c]{27.774}          & \makecell*[c]{2.760}           & \makecell*[c]{98.1\%}         & \makecell*[c]{67.2\%}         \\ 
            \makecell*[c]{500}            & \makecell*[c]{12606(Dense)}     & \makecell*[c]{60}                & \makecell*[c]{66.590}          & \makecell*[c]{88.406}          & \makecell*[c]{1.720}           & \makecell*[c]{99.0\%}         & \makecell*[c]{63.2\%}         \\ 
            \makecell*[c]{2000}           & \makecell*[c]{15037(Sparse)}       & \makecell*[c]{15}                & \makecell*[c]{30.428}          & \makecell*[c]{20.664}          & \makecell*[c]{3.391}           & \makecell*[c]{99.5\%}         & \makecell*[c]{98.7\%}         \\ 
            \makecell*[c]{2000}           & \makecell*[c]{15037(Sparse)}       & \makecell*[c]{60}                & \makecell*[c]{101.564}         & \makecell*[c]{71.571}          & \makecell*[c]{14.048}          & \makecell*[c]{99.4\%}         & \makecell*[c]{98.4\%}         \\ 
            \makecell*[c]{2000}           & \makecell*[c]{199452(Dense)}    & \makecell*[c]{15}                & \makecell*[c]{38.593}          & \makecell*[c]{41.987}          & \makecell*[c]{1.706}           & \makecell*[c]{100\%}          & \makecell*[c]{82.5\%}         \\ 
            \makecell*[c]{2000}           & \makecell*[c]{199452(Dense)}    & \makecell*[c]{60}                & \makecell*[c]{68.729}          & \makecell*[c]{104.784}         & \makecell*[c]{1.356}           & \makecell*[c]{100\%}          & \makecell*[c]{57.9\%}         \\ 
       
            \bottomrule
        \end{tabular}}
\end{table*}

The connectivity or structure of subgraphs is undeniably a crucial factor influencing network capacity, as it determines the resource usage of each subgraph. There are two main factors that impact subgraph connectivity: the scale of a single sample and network connectivity. Table \ref{tab:2} demonstrates that a larger scale of a single sample and better network connectivity will reduce network capacity. This observation is intuitive for the former but counterintuitive for the latter. However, when the scale of a single sample node is 0 or network connectivity is extremely poor, the network capacity tends to be 0. This suggests that the relationship between network capacity and subgraph connectivity is not linear.

\begin{figure}[h]
  \centering
  \includegraphics[width=\linewidth]{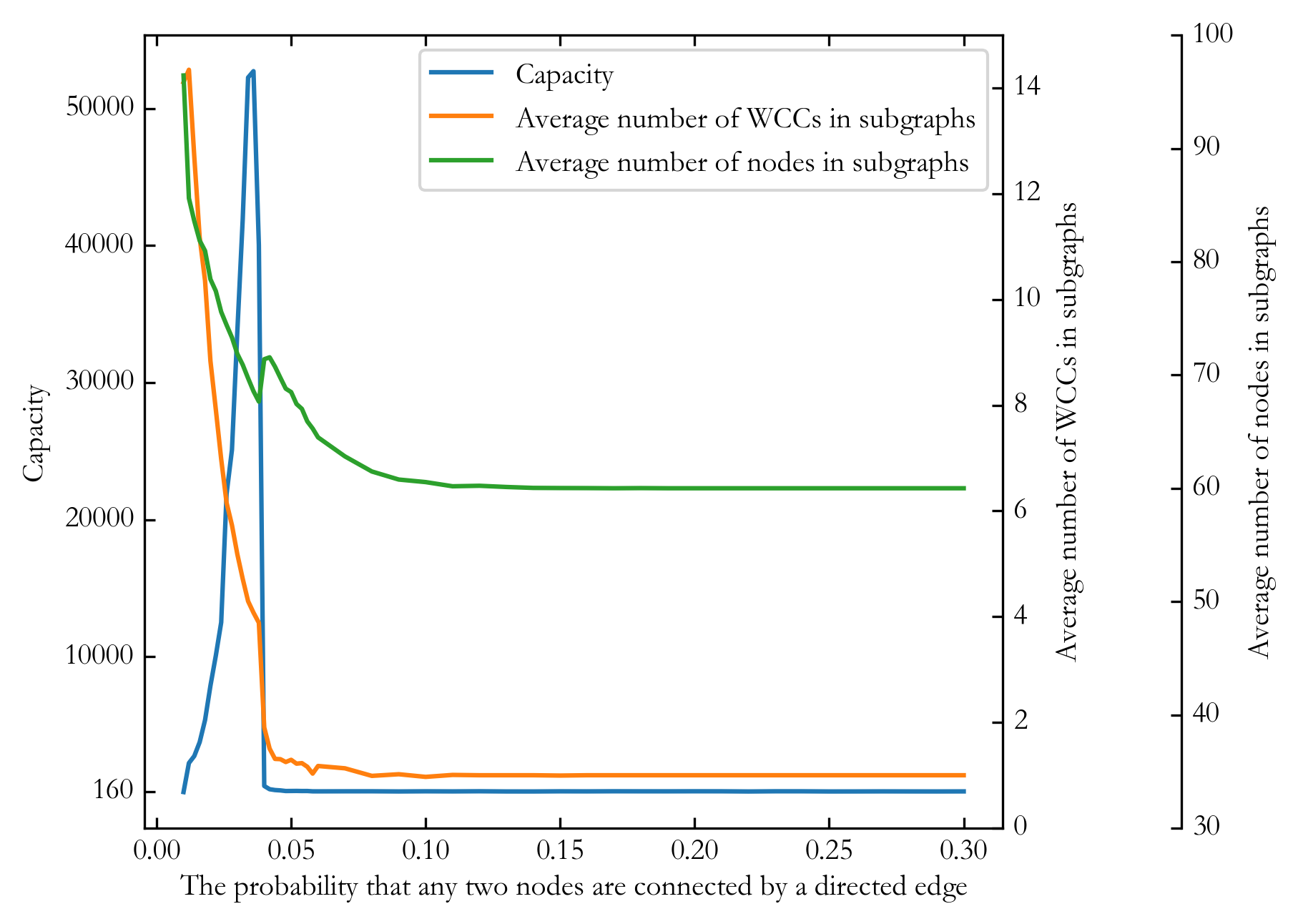}
  \caption{The relationship between the number of edges and network capacity and subgraph structure}
  \label{fig:7}
\end{figure}

Figure \ref{fig:7} shows the changes in network capacity and subgraph structure as the number of edges in a network increases. The node size of the network is 500, and the single sample scale is 60. It can be observed that the network capacity first rises and then declines, eventually stabilizing near the theoretical capacity value in the simple case, which is $\frac{nK}{s+c}$. During the stage of network capacity growth, both the average number of WCCs and the average number of nodes in the subgraph decrease, suggesting that the subgraph progressively transitions from "dispersed" to "clustered." Subsequently, there is a sharp decline in network capacity, and the average number of WCCs in the subgraph also drops dramatically. This indicates that the network connectivity has reached a critical point, with almost all nodes in the subgraph belonging to the same WCC. As a result, the "agglomeration effect" emerges. It means most initial nodes can connect directly without passing through other communicating nodes. When the network capacity reaches its peak, the average number of nodes in the subgraphs is close to the scale of a single sample, and the number of WCCs is slightly above 1.

\begin{figure}[h]
  \centering
  \includegraphics[width=\linewidth]{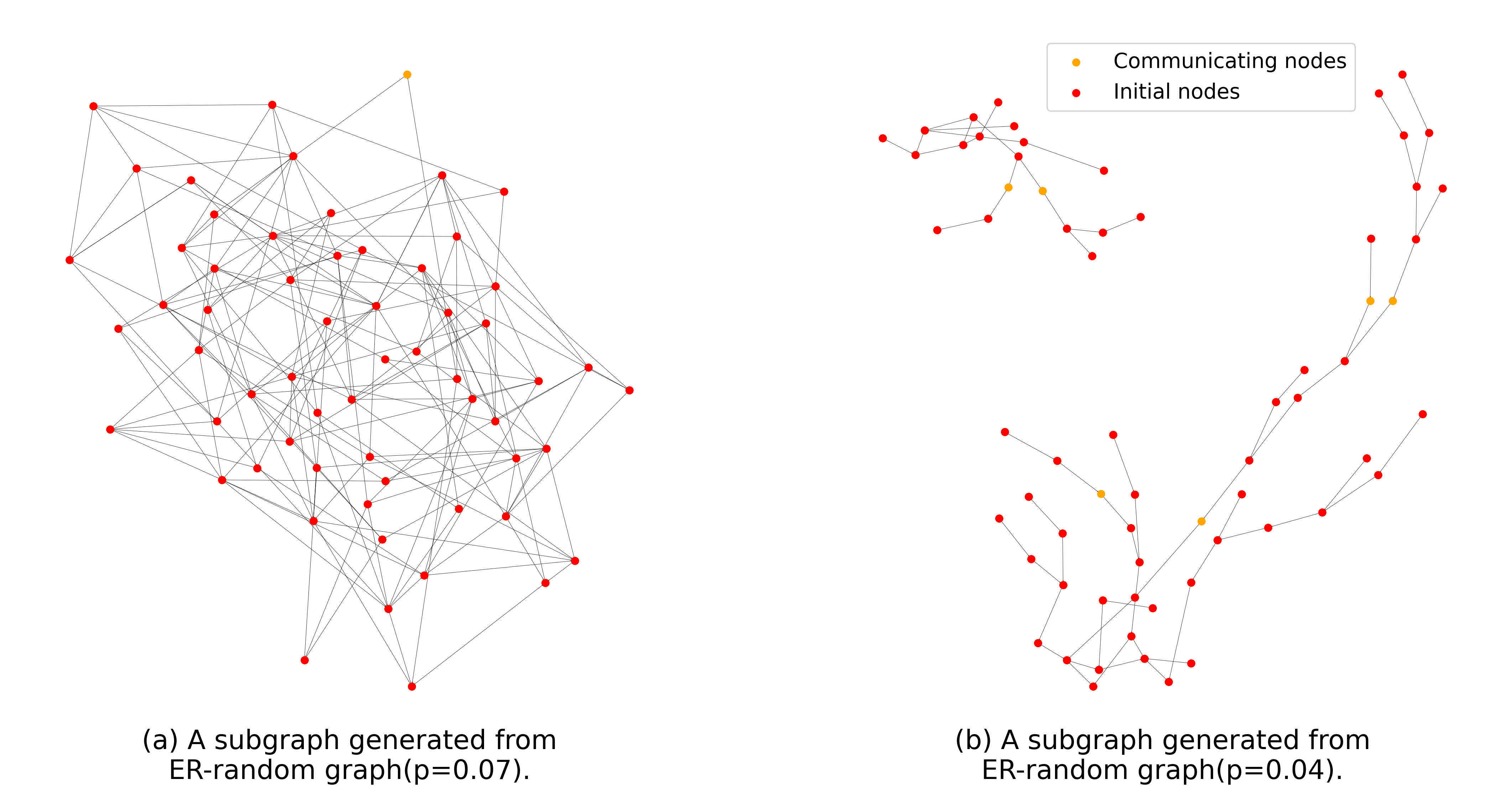}
  \caption{Storage examples under different network connectivity. (a) The sample-generated subgraph in an ER random graph with p=0.07. (b) The sample-generated subgraph in an ER random graph with p=0.04.}
  \label{fig:8}
\end{figure}

Erdős and Rényi \cite{erdosEvolutionRandomGraphs2006} demonstrated that when $p>\frac{(1+\epsilon) \ln n}{n}$, the ER random graph $G(n,p)$ is almost always connected. To ensure that the subgraphs generated by the samples have a high probability of only 1 WCC, it needs to guarantee that $p>\frac{(1+\epsilon) \ln s}{s}$, where $s$ represents the scale of a single sample, which is 60. Let's take $p=\frac{\ln s}{s}\approx 0.07$. The generated subgraph in this scenario is shown in Figure \ref{fig:8}a. If we take $p=0.04$, corresponding to the $p$ when the network capacity reaches its peak, the generated subgraph is shown in Figure \ref{fig:8}b. It can be found that the essence of large capacity is actually the permutation and combination of multiple WCCs. When $p$ is slightly less than $\frac{\ln s}{s}$, the subgraphs generated by the samples are composed of a small number of WCCs. Assuming the subgraph is evenly divided into $t$ WCCs, the size of each WCC is $\frac{s+c}{t}$. The network capacity can be perceived as selecting $t$ WCCs from all possible ones. This is essentially a Uniform disordered grouping problem. The calculation result is shown in formula \ref{eq:1}. Although the actual capacity is significantly smaller than this value, it still demonstrates the huge storage potential of the network.

\begin{equation}
    \begin{aligned}
        T &=& \frac{\tbinom{n}{\frac{s+c}{t}}  \tbinom{n - \frac{s+c}{t}}{\frac{s+c}{t}} ... \tbinom{n - (t-1)\frac{s+c}{t}}{\frac{s+c}{t}}}{t!}  \\
        &=& \frac{n!}{(t!)^{\frac{s+c+t}{t}} (n-s-c)!} 
    \end{aligned}
    \label{eq:1}
\end{equation}

\begin{figure}[h]
  \centering
  \includegraphics[width=\linewidth]{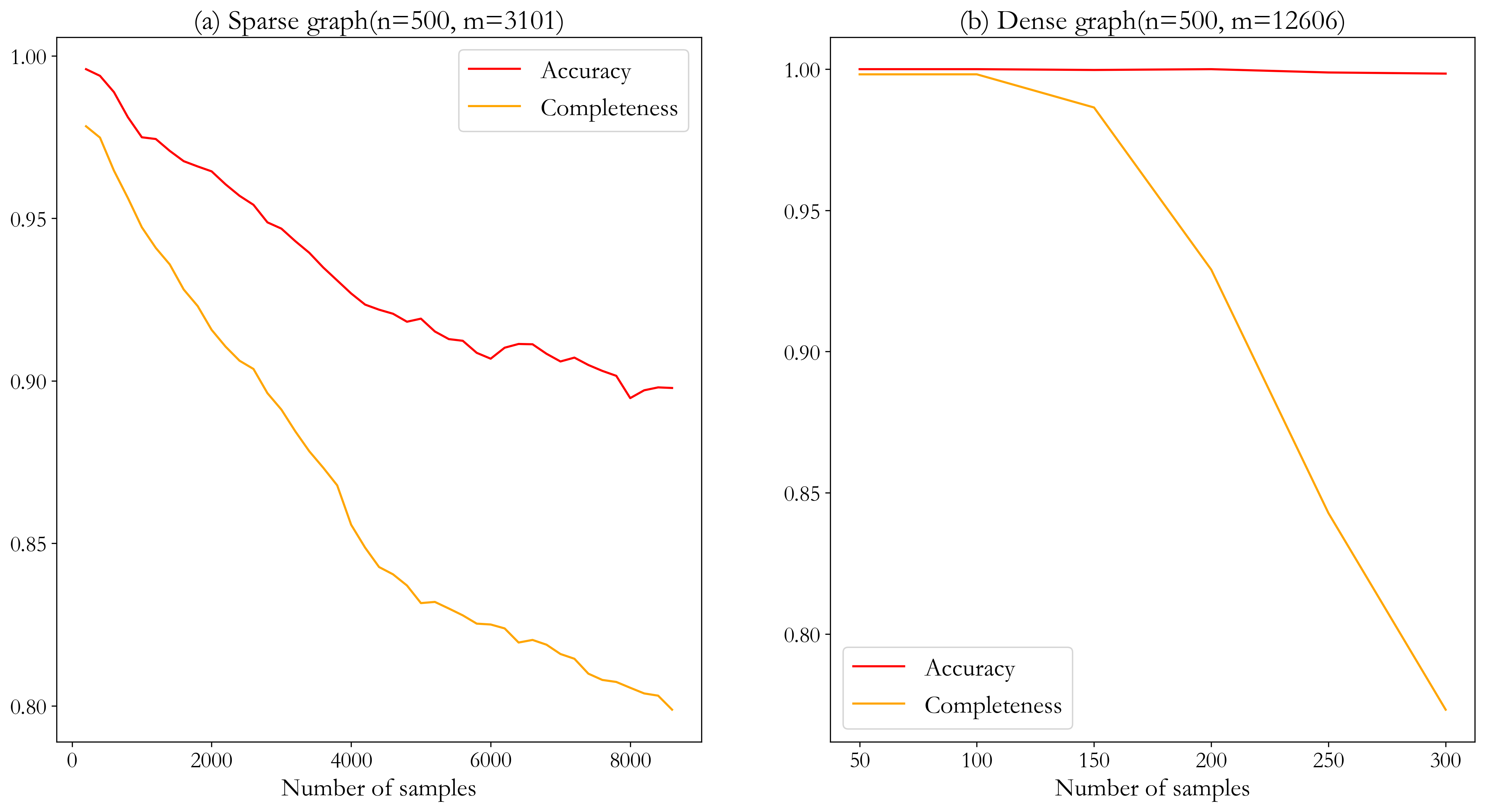}
  \caption{Comparison of capacity between sparse and dense graphs. (a) Results in sparse graph. (b) Results in dense graph.}
  \label{fig:9}
\end{figure}

Figure \ref{fig:9} demonstrates the capacity difference between a sparse graph and a dense graph, both with the same number of nodes. It can be observed that in the sparse graph, the average completeness of sample retrieval drops below 80\% when the number of stored samples exceeds 8000. In contrast, for the dense graph, the average completeness of sample retrieval declines below 80\% when the number of stored samples approaches 300. This capacity difference between the two networks further confirms that the arrangement and combination of WCCs are the essences of large capacity. Although the dense graph has more connections, its displayed capacity is not directly proportional to the number of resources owned by the network. Conversely, the sparse graph has only a small number of connections, but the network capacity achieved by the arrangement and combination of multiple connected subgraphs is several times that of the dense graph. This indicates that a sparse connection is a more reasonable mode, which can effectively save resources and obtain a larger network capacity. Moreover, the biological neuron network of the human brain also follows a sparse connection mode, implying that sparse connections are efficient and maximize the use of resources.

\subsection{Fault tolerance testing}

The fault tolerance testing primarily explores the effect of sample retrieval when the input is incomplete or has noise. The network used in the experiment is an ER random graph with 500 nodes and 3101 (sparse graph) and 12606 (dense graph) edges, respectively. The experiment first stores 1000 samples in the network, then modifies the sample inputs during the retrieval process. Finally, compare how the average accuracy and completeness change when the sample input is incomplete or has noise. There are three categories of modifications to inputs:

1.	Removing a part of the original sample input to explore the impact of incomplete inputs on sample retrieval performance.

2.	Adding extra noisy nodes to the original sample to investigate the impact of noise on retrieval.

3.	Removing part of the original sample input and replacing it with an equal number of noisy nodes to examine the impact of sample retrieval in this mixed scenario.

\begin{figure}[h]
  \centering
  \includegraphics[width=\linewidth]{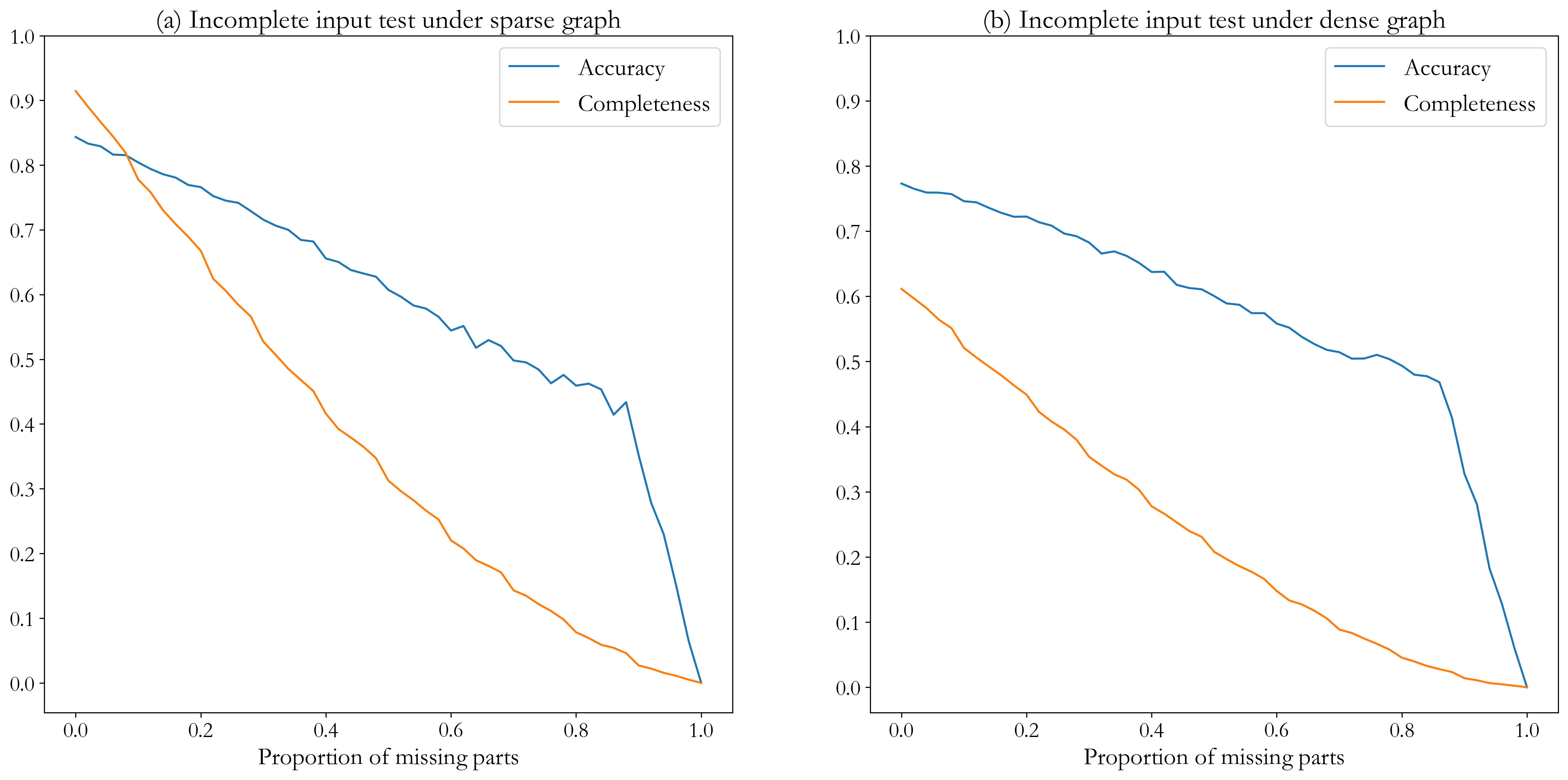}
  \caption{The impact of incomplete sample input on sample retrieval. (a) Test results in a sparse graph. (b) Test results in a dense graph.}
  \label{fig:incmp}
\end{figure}

Figure \ref{fig:incmp} shows that as more sample inputs are missing, the accuracy and completeness of retrieval decrease to varying degrees in both sparse and dense graphs. The change trends of the two networks are generally similar, and the decline in accuracy is relatively gentle. When the proportion of missing parts reaches 80\% of the input, the rate of accuracy decline increases significantly. Compared to Figure \ref{fig:incmp}b, Figure \ref{fig:incmp}a has higher completeness and accuracy when the proportion of missing inputs is between 0.0 and 0.1. This is because the network capacity of the dense graph is small, making it difficult to achieve high reading accuracy and completeness after storing 1000 samples. The overall decline rate of completeness is greater than that of accuracy, indicating that the erroneous content obtained during retrieval does not increase as the proportion of missing inputs increases. This suggests that the algorithm is relatively reliable when facing missing sample inputs, although the rapid decline in completeness represents a significant amount of correct content that cannot be read. However, even when the proportion of missing inputs is as high as 80\%, the retrieval accuracy can still be maintained at around 40\% to 50\%, which means that even if there are numerous missing inputs, almost half of the read content is correct and reliable.

\begin{figure}[h]
  \centering
  \includegraphics[width=\linewidth]{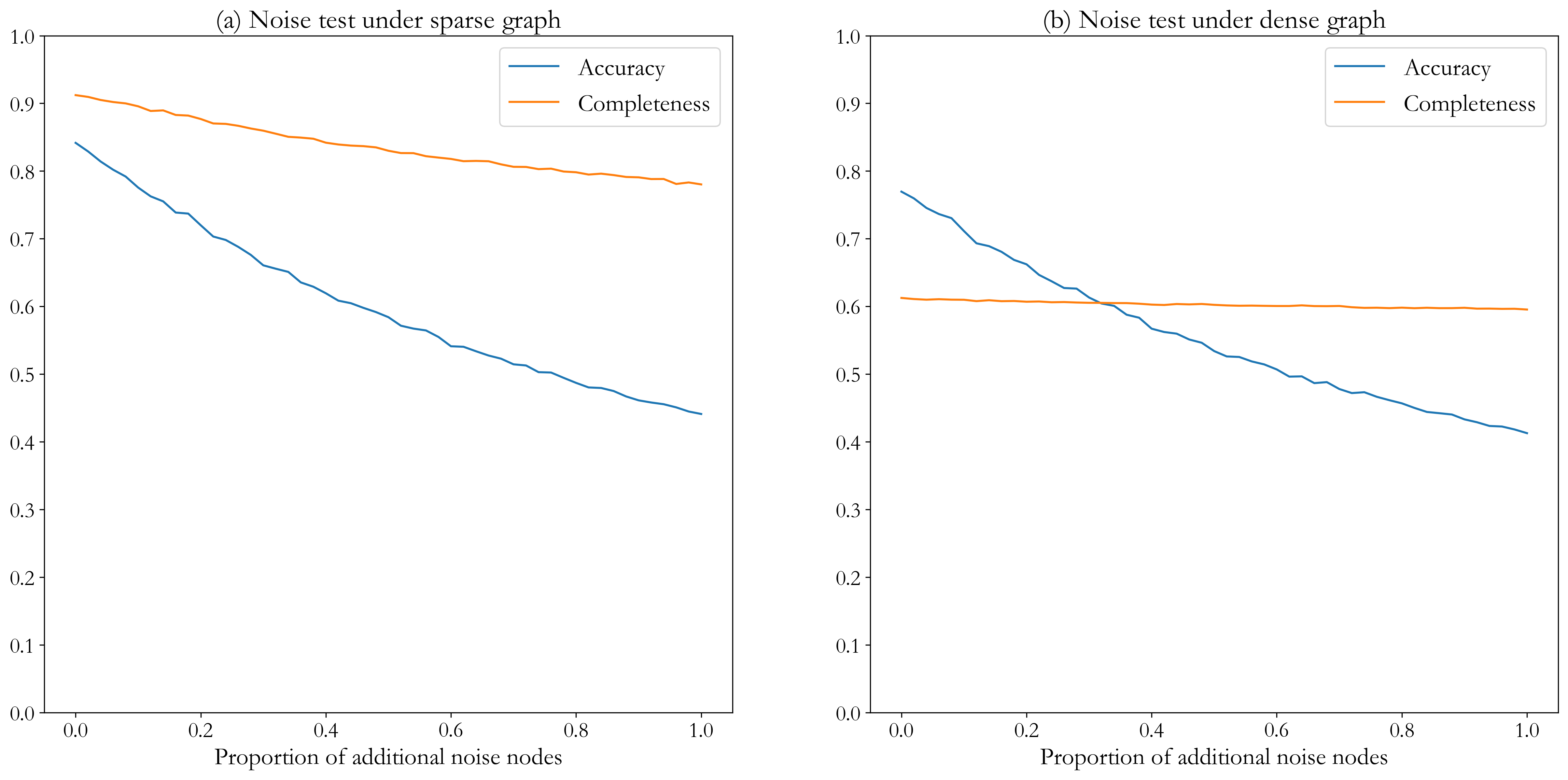}
  \caption{The impact of noisy sample input on sample retrieval. (a) Test results in a sparse graph. (b) Test results in a dense graph.}
  \label{fig:wrong}
\end{figure}

Figure \ref{fig:wrong} shows the impact of increasing noise nodes. It can be observed that these additional noise nodes have relatively little effect on the accuracy and completeness of sample retrieval. The decline rate of completeness is lower than that of accuracy because adding noise nodes generally do not directly disrupt the original subgraph structure but makes the final subgraph larger. This demonstrates that the network has a relatively strong resistance to noise and is sensitive to the absence of sample inputs.

\begin{figure}[h]
  \centering
  \includegraphics[width=\linewidth]{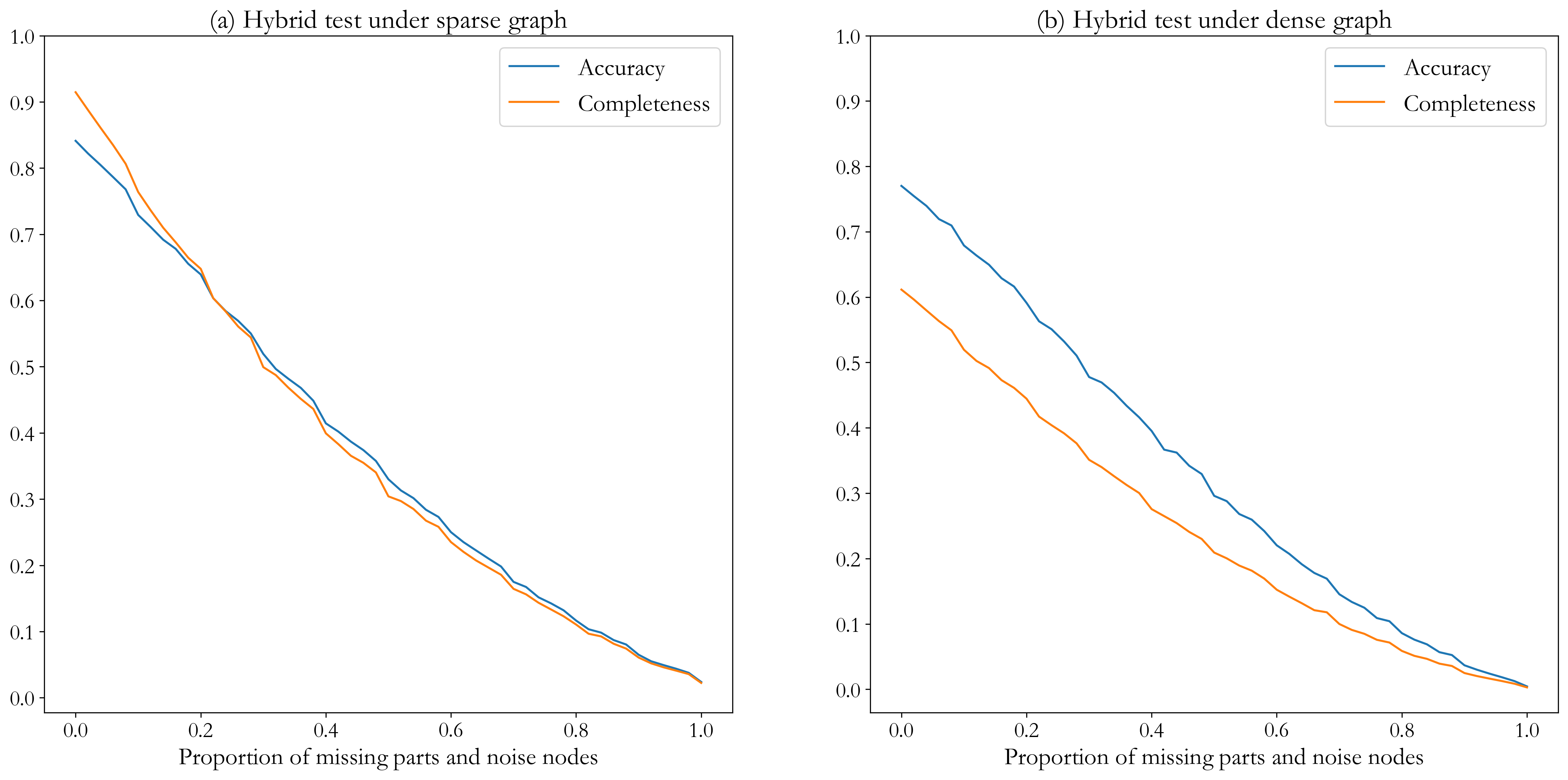}
  \caption{The impact of both incomplete and contained noise nodes in the sample input on sample retrieval. (a) Test results in a sparse graph. (b) Test results in a dense graph.}
  \label{fig:maw}
\end{figure}

Figure \ref{fig:maw} presents the performance of both incomplete and contained noise nodes in the sample input under sparse and dense graphs, respectively. It can be seen that the decline rate of accuracy and completeness, in this case, is the fastest, indicating that the impact of missing inputs and the effect of noise nodes can be superimposed.

\subsection{Robustness testing}

It is known that neurons in the biological brain may experience various functional failures. How does this affect memory? This section primarily examines how the performance of sample retrieval changes when the network is damaged to different degrees. The test includes two main aspects: damage to some nodes and damage to some directed edges. The experiment first stores 1000 samples in the network, then deletes a certain proportion of nodes or directed edges and subsequently attempts to retrieve these samples while recording average accuracy and completeness changes. The network used in the experiment is an ER random graph with 500 nodes and 3101 edges (sparse graph) or 12606 edges (dense graph).

After a node or a directed edge is deleted, the activation traces recorded in the node's internal index table will be affected. Assume that an index table contains two items: {A,B,C} → {X,Y,Z} and {B,C,D} → {U,X,Z}. If nodes A, D, and Z are deleted, does it need to delete the corresponding node in the activation trace recorded in its index table? If deleted, then these two items will become: {B,C} → {X,Y} and {B,C} → {U,X}. It can be observed that the input parts of these two items are the same, so addressing the different output parts is a challenge. Usually, during the initial period after network damage, nodes are hard to respond, and at this time, the original traces stored in the node index table will not change. As the damage duration increases, nodes may gradually make corresponding adaptive adjustments to the damaged network. Given the above two different situations, this paper proposes four restoration schemes, as shown in Table \ref{tab:3}, and compares these four solutions.

\begin{table}[]
    \caption{Restoration schemes for the index table after damage}
    \label{tab:3}
        \begin{tabular}{cc}
            \toprule

            \makecell*[c]{\textbf{Restoration schemes}}       &  \makecell*[c]{\textbf{The modified content of} \\ \textbf{the index table}}                                                  \\ 

            \midrule

            \makecell*[c]{Maintain the original trace}  &  \makecell*[c]{\{A,B,C\} $\rightarrow$ \{X,Y,Z\}, \\ \{B,C,D\} $\rightarrow$ \{U,X,Z\}} \\ 
            \makecell*[c]{Take the union of the outputs}            &  \makecell*[c]{\{B,C\} $\rightarrow$ \{X,Y,U\}     }                              \\ 
            \makecell*[c]{Take the intersection of the outputs}            &  \makecell*[c]{\{B,C\} $\rightarrow$ \{X\}   }                                      \\ 
            \makecell*[c]{Take the output with the highest \\ occurrence frequency}          &  \makecell*[c]{\{B,C\} $\rightarrow$ \{X,Y\} or   \{U,X\}  }                         \\ 
     
            \bottomrule
        \end{tabular}
\end{table}

\begin{figure}[h]
  \centering
  \includegraphics[width=\linewidth]{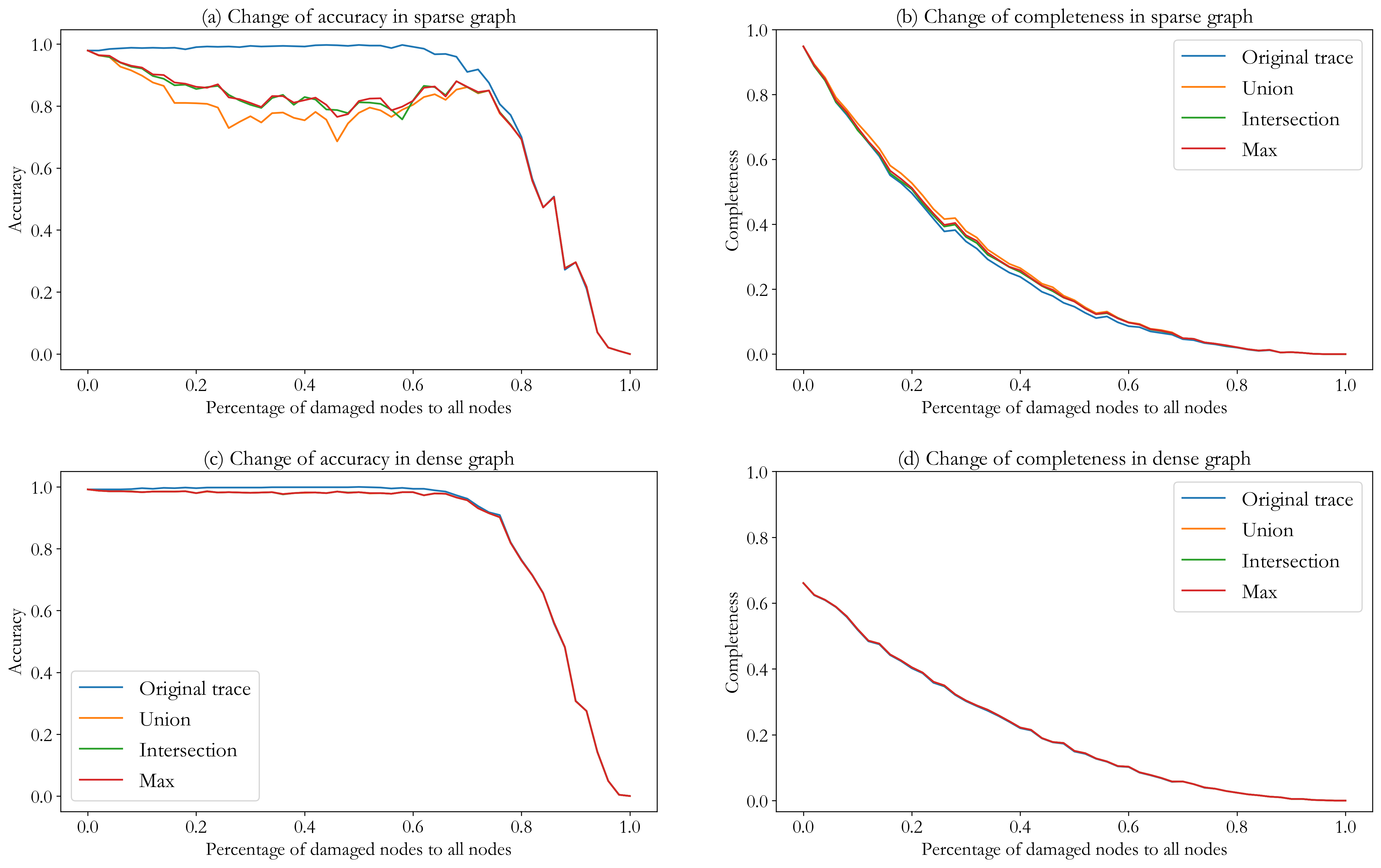}
  \caption{The impact of partial node damage on sample retrieval. (a) The change in accuracy in a sparse graph. (b) The change in completeness in a sparse graph. (c) The change in accuracy in a dense graph. (d) The change in completeness in a dense graph.}
  \label{fig:dp-en-new}
\end{figure}

Figure \ref{fig:dp-en-new} demonstrates the impact of partial node damage on sample retrieval performance. Figure \ref{fig:dp-en-new}a and Figure \ref{fig:dp-en-new}b respectively display the changes in average accuracy and completeness of sample retrieval in the sparse graph for the four restoration schemes. In terms of average accuracy, the scheme that maintains the original traces performs the best, the scheme that takes the union performs the worst, and the other two schemes exhibit similar performance. Conversely, in terms of average completeness, the results are reversed. The scheme that takes the union performs the best, while the one that maintains the original traces performs the worst. This is because the union-taking scheme increases the number of activated nodes, which includes both incorrect and correct nodes. The former leads to a decrease in accuracy, while the latter leads to an increase in completeness. Figure \ref{fig:dp-en-new}c and Figure \ref{fig:dp-en-new}d present the results on the dense graph, revealing that when the network has a large number of edges, the differences between the four restoration schemes progressively diminish. This occurs because when network connectivity is high, the number of communication nodes in the subgraph generated by the sample is small, with most initial nodes being directly connected. Consequently, the accuracy remains consistently high. The frequency at which each node is shared by different samples is also reduced, so when a node is deleted, the number of samples it affects decreases, making the differences between the four solutions less noticeable. 

\begin{figure}[h]
  \centering
  \includegraphics[width=\linewidth]{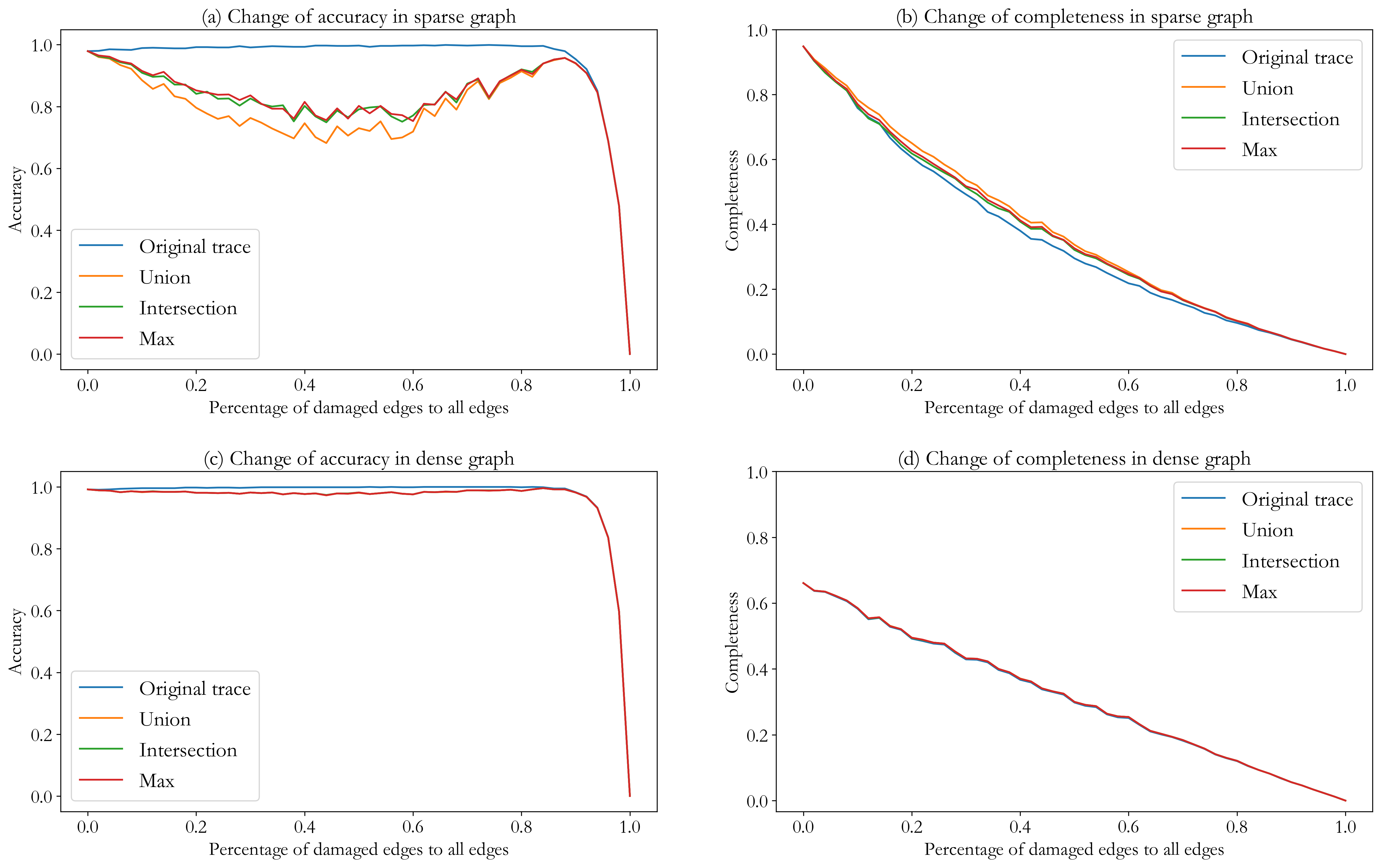}
  \caption{The impact of partial directed edge damage on sample retrieval. (a) The change in accuracy in a sparse graph. (b) The change in completeness in a sparse graph. (c) The change in accuracy in a dense graph. (d) The change in completeness in a dense graph.}
  \label{fig:de-en-new}
\end{figure}

Figure \ref{fig:de-en-new} shows the impact of partial directed edge damage on sample retrieval. Since a node only has a local view, it can receive and transmit the information of neighboring nodes solely through its fan-in and fan-out edges. Node damage can be understood as the interruption of all fan-in and fan-out connections, so the impact on neighboring nodes is essentially the same, whether node damage or directed edge damage. Consequently, the same restoration schemes can be used. Figure \ref{fig:de-en-new}a and Figure \ref{fig:de-en-new}b respectively display the changes in average accuracy and completeness of sample retrieval for the four restoration schemes in the sparse graph. Their trends are almost consistent with Figure \ref{fig:dp-en-new}a and Figure \ref{fig:dp-en-new}b. Regarding accuracy, the notable difference between the two is that in the interval [0.8,1.0], Figure \ref{fig:de-en-new}a maintains relatively high accuracy. Regarding completeness, the curve of Figure \ref{fig:de-en-new}b is comparatively flat. Figure \ref{fig:de-en-new}c and Figure \ref{fig:de-en-new}d are the results of dense graphs. The results are also consistent with those of Figure \ref{fig:dp-en-new}c and Figure \ref{fig:dp-en-new}d. The difference is the same as that observed in the sparse graph, which indicates that the network is significantly more tolerant of directed edge damage than node damage, as nodes hold information while edges do not.

\subsection{Performance testing on different classical network structures}

The information storage and retrieval algorithm proposed in this paper is closely related to the network structure. Firstly, the algorithm utilizes the subgraph structure as the information storage carrier, and secondly, the subgraph formation depends on stimulus propagation. Both of these characteristics emphasize the importance of the network structure for the algorithm. Therefore, different network structure characteristics are key factors affecting the algorithm's performance.

\begin{figure}[h]
  \centering
  \includegraphics[width=\linewidth]{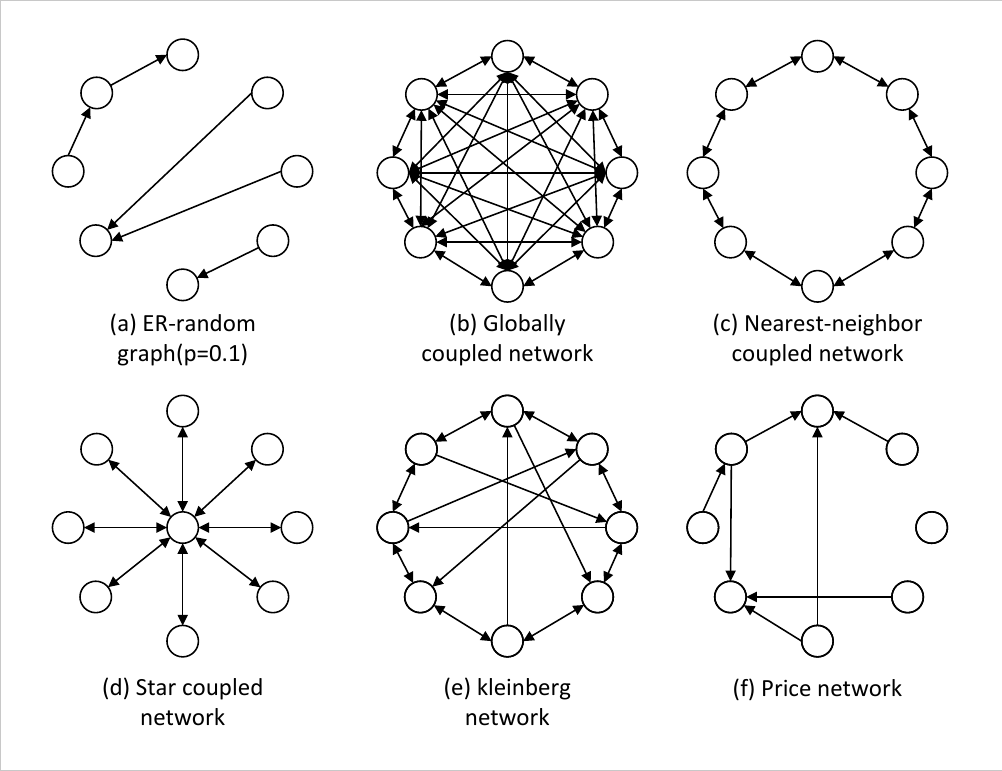}
  \caption{Six classic network structures. (a) ER random graph with p=0.1. (b) Globally coupled network. (c) Nearest-neighbor coupled network with L=1. (d) Star coupled network. (e) Kleinberg network. (f) Price network.}
  \label{fig:graphs}
\end{figure}

Figure \ref{fig:graphs} showcases six classic network structures. Figure \ref{fig:graphs}a is the ER graph with $p=0.1$. Figure \ref{fig:graphs}b represents a globally coupled network, also known as a fully connected network. Figure \ref{fig:graphs}c shows the nearest-neighbor coupled network, characterized by $N$ nodes arranged in a ring, with each node establishing connections to its left and right $L$ neighbors, respectively. Figure \ref{fig:graphs}d illustrates a star coupled network, featuring a central node to which all other nodes are connected. This characteristic causes any path between two points in the network to include the central node, creating a bottleneck for the entire network capacity. Figure \ref{fig:graphs}e presents a one-dimensional Kleinberg network \cite{easleyNetworksCrowdsMarkets2010}, a small-world network \cite{wattsCollectiveDynamicsSmallworld1998}. The network is constructed by adding a few random edges to the nearest-neighbor coupled network. Figure \ref{fig:graphs}f displays the Price network \cite{priceNetworksScientificPapers1965}, which is a scale-free network. The network's generation relies on the preferential attachment mechanism, where newly added nodes are more likely to connect to nodes with higher degrees. Since each newly added directed edge point from the new node to the old node, there are no loops in the network, leading to a significant decrease in network connectivity and capacity.

Evaluation parameters for different network structures typically include average path length and clustering coefficient.

Average Path Length: Defined as the average of the shortest path lengths between any two points in the network. The default shortest path length is usually positive infinity if the two nodes are disconnected. This special case is common in directed graphs. To prevent the calculation result from being positive infinity, this paper uses the harmonic mean \cite{wangxiaofanWangluokexuedaolun2012b} of the distance between any two nodes in the network to represent the average path length.

$N$ represents the number of network nodes, and $d(i, j)$ represents the shortest distance between node $i$ and node $j$. $GE$ represents network communication efficiency, with its essential idea being that the closer the node path distance in the network, the higher the communication efficiency. The average path length calculated by the formula \ref{eq:2} \cite{wangxiaofanWangluokexuedaolun2012b} solves the problem of the value being positive infinity when the network is disconnected. Therefore, it is more suitable for evaluating directed graph network structure.

\begin{equation}
    L=\frac{1}{GE}, GE = \frac{1}{N(N-1)} \sum_{i\geq j}\frac{1}{d(i,j)} \label{eq:2}
\end{equation}

Clustering coefficient: This metric is used to measure whether the nodes in the network exhibit aggregation characteristics. This paper adopts the calculation method of the clustering coefficient in directed graphs proposed by Fagiolo \cite{fagioloClusteringComplexDirected2007a}.

\begin{figure}[h]
  \centering
  \includegraphics[width=\linewidth]{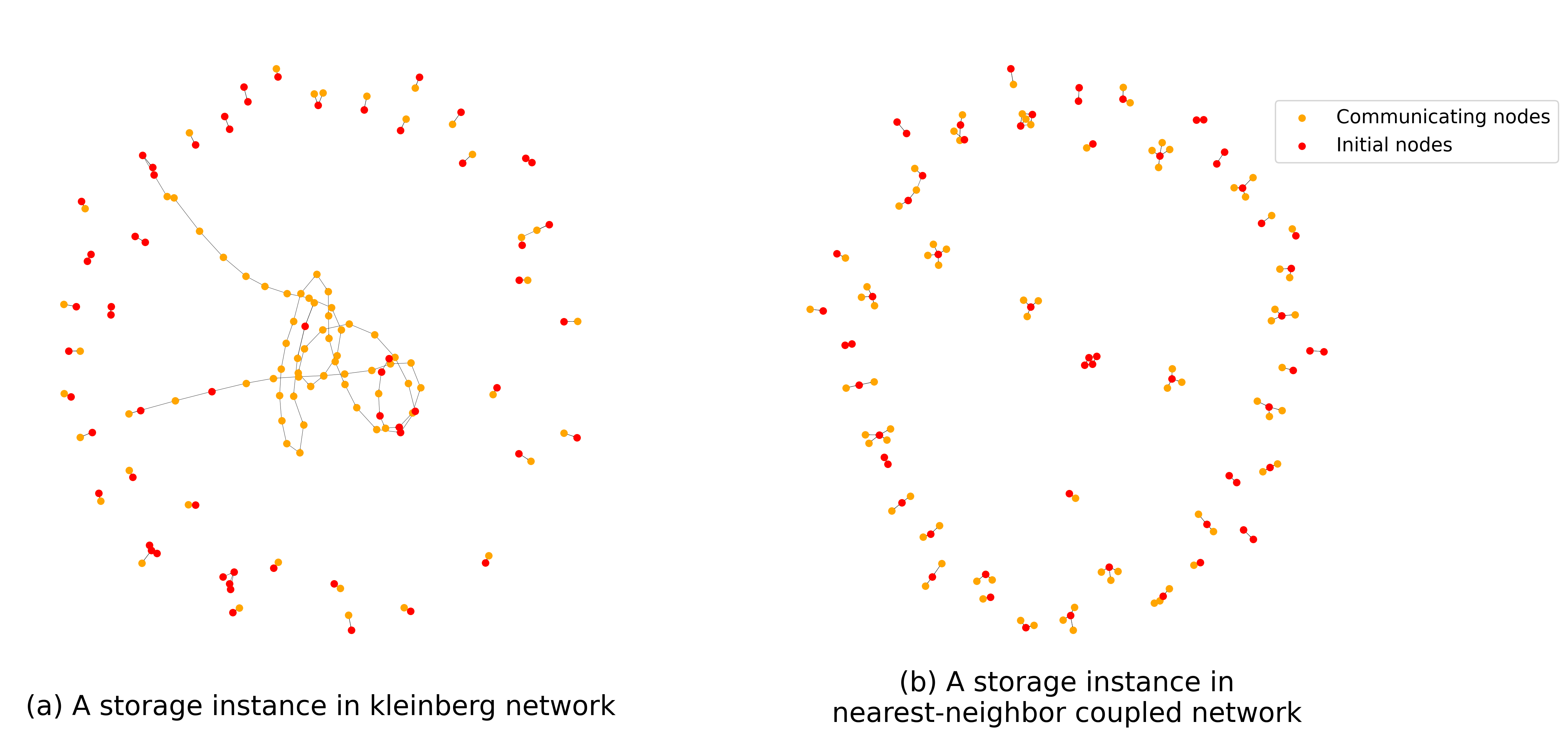}
  \caption{Network storage example (a) Kleinberg network storage example. (b) Nearest-neighbor coupled network storage example.}
  \label{fig:kelin}
\end{figure}

Table \ref{tab:4} displays the test results of six network models with different structures but similar scales. The number of nodes in all test networks is 1000, and the single sample scale is 60. The ER random graph exhibits a small clustering coefficient and average path length, while the nearest-neighbor coupled network has a large clustering coefficient and average path length. The Kleinberg directed small-world network has a large clustering coefficient and a small average path length. Kleinberg's directed small-world and nearest-neighbor coupled network demonstrate relatively excellent network capacity among these six network types. Figure \ref{fig:kelin} illustrates examples of sample storage corresponding to the two networks. A common feature observed in both networks is the presence of numerous small WCCs. As previously mentioned in the network capacity analysis, the essence of large network capacity lies in the arrangement and combination of WCCs. Since both networks have relatively high clustering coefficients, it is quite easy to form small components locally. Each subgraph can be considered a combination of several small components, resulting in a higher network capacity. The reason for the higher capacity of the Kleinberg network is that, due to its lower average path length, it is easier to form some large WCCs. These large WCCs not only exhibit higher distinguishability but also have a higher resource reuse rate for their nodes, thus positively impacting the improvement of network capacity. In contrast, these characteristics are not present in the ER random graph. Due to its low clustering coefficient and average path length, most weakly connected components formed by the ER random graph are large in scale. This is the difference in capacity caused by different network structures.

\begin{table*}[]
    \caption{Comparison of classic network models}
    \label{tab:4}
    \resizebox{\textwidth}{!}{
        \begin{tabular}{cccccccc}
            \toprule

            \makecell*[c]{\textbf{Network} \\ \textbf{Type}}    & \makecell*[c]{\textbf{Number} \\ \textbf{of edges}} & \makecell*[c]{\textbf{Clustering} \\ \textbf{coefficient}} & \makecell*[c]{\textbf{Average} \\ \textbf{path length}} & \makecell*[c]{\textbf{Maximum} \\ \textbf{reliable capacity}} & \makecell*[c]{\textbf{Subgraph average} \\ \textbf{number of nodes}} & \makecell*[c]{\textbf{Subgraph average} \\ \textbf{number of edges}}  & \makecell*[c]{\textbf{Subgraph average} \\ \textbf{number of WCCs}} \\ 
            
            \midrule

            \makecell*[c]{ER random \\ graph}            & \makecell*[c]{6070}        & \makecell*[c]{0.006}           & \makecell*[c]{3.797}           & \makecell*[c]{85}              & \makecell*[c]{128.139}         & \makecell*[c]{116.708}                   & \makecell*[c]{12.044}            \\ 
            \makecell*[c]{Globally coupled \\ network}           & \makecell*[c]{999000}      & \makecell*[c]{1.000}           & \makecell*[c]{1.000}           & \makecell*[c]{341}             & \makecell*[c]{60.000}          & \makecell*[c]{180.000}                     & \makecell*[c]{1.000}             \\ 
            \makecell*[c]{Nearest-neighbor \\ coupled}          & \makecell*[c]{6000}        & \makecell*[c]{0.600}           & \makecell*[c]{29.235}          & \makecell*[c]{576}             & \makecell*[c]{167.234}         & \makecell*[c]{241.087}                    & \makecell*[c]{18.768}            \\ 
            \makecell*[c]{Star coupled \\ network}           & \makecell*[c]{1998}        & \makecell*[c]{0.000}           & \makecell*[c]{1.996}           & \makecell*[c]{21}              & \makecell*[c]{60.900}          & \makecell*[c]{62.900}                     & \makecell*[c]{1.000}             \\ 
            \makecell*[c]{Price \\ network}     & \makecell*[c]{5978}        & \makecell*[c]{0.170}           & \makecell*[c]{85.116}          & \makecell*[c]{0}               & \makecell*[c]{0}               & \makecell*[c]{0}                                & \makecell*[c]{0}                 \\ 
            \makecell*[c]{Kleinberg \\ network} & \makecell*[c]{6995}        & \makecell*[c]{0.440}           & \makecell*[c]{4.613}           & \makecell*[c]{6144}            & \makecell*[c]{98.320}          & \makecell*[c]{98.180}                     & \makecell*[c]{24.084}            \\ 
            
            \bottomrule
        \end{tabular}}
\end{table*}

\subsection{Conclusion}

In this paper, we employ subgraphs as physical carriers for information storage and leverage nodes' autonomous adaptive learning behavior to achieve a large-capacity and stable directed graph storage model. The individual nodes' learning behavior does not need a global view, meaning that the tiny algorithms operating within each node do not work under strong central control and are entirely decentralized. Both the learning behavior and the supporting hardware resources are fine-grained and distributed and can, in theory, be highly parallel in physical implementation.

The storage capacity of the network depends on factors such as connectivity and network structure. The dense graph has better connectivity, the subgraphs generated by the samples are usually gathered together, and the communication nodes are rarely used. The measured capacity at this time is low, approaching the theoretical capacity limit that disallows resource reuse. Sparse graphs exhibit poor connectivity, and the sample-generated subgraphs are generally more dispersed, often consisting of several weakly connected components. In this case, the sample-generated subgraphs can be viewed as a permutation of connected components, significantly increasing the network capacity. Tests have shown that a sparse random directed graph with 500 nodes and 3101 edges can store nearly 8000 memory samples with over 80\% accuracy and completeness. In contrast, a dense graph with 500 nodes and 12606 edges can only store around 300 memory samples.

Sparse graphs have fewer resources than dense graphs, but the actual number of samples they can store is tens of times more than dense graphs. It demonstrates that resource abundance is not the sole factor determining network capacity. The network's structural properties, such as connectivity, clustering coefficient, and average path length, are also crucial. Biological neuronal networks exhibit sparse connections and show large capacity and low power consumption characteristics. To some extent, this paper also provides a possible explanation for how biological neuronal networks can achieve memory functions.

\bibliographystyle{IEEEtran}
\bibliography{main}


 




\vfill

\end{document}